\documentclass[12pt,preprint]{emulateapj}
\usepackage{natbib,amssymb}
\usepackage{amsmath}

\shorttitle{SONYC IX: Cha-I \& Lupus~3}
\shortauthors{Mu\v{z}i\'c et al.}

\begin{document}
\bibliographystyle{apj}

\newcommand{\solm}{M$_{\odot}$}
\newcommand{\soll}{L$_{\odot}$}
\newcommand{\solr}{R$_{\odot}$}

\title{Substellar Objects in Nearby Young Clusters (SONYC) IX: The planetary-mass domain of Chamaeleon-I and updated mass function in Lupus-3 \altaffilmark{*}}

\author{Koraljka Mu\v{z}i\'c\altaffilmark{1,**}, Alexander Scholz\altaffilmark{2}, Vincent C. Geers\altaffilmark{3,4}, Ray Jayawardhana\altaffilmark{5}}

\email{kmuzic@eso.org}

\altaffiltext{1}{European Southern Observatory, Alonso de C\'ordova 3107, Casilla 19, Santiago, 19001, Chile}
\altaffiltext{2}{School of Physics \& Astronomy, St. Andrews University, North Haugh, St Andrews KY16 9SS, United Kingdom}
\altaffiltext{3}{UK Astronomy Technology Centre, Royal Observatory Edinburgh, Blackford Hill, Edinburgh, EH9 3HJ, United Kingdom}
\altaffiltext{4}{School of Cosmic Physics, Dublin Institute for Advanced Studies, 31 Fitzwilliam Place, Dublin 2, Ireland}
\altaffiltext{5}{Faculty of Science, York University, 355 Lumbers Building, 4700 Keele Street, Toronto, ON M3J 1P2, Canada}
\altaffiltext{*}{Based on observations collected at the European Southern Observatory under the program 093.C-0050(A).}
\altaffiltext{**}{Present address: Nucleo de Astronom\'ia, Facultad de Ingenier\'ia, Universidad Diego Portales, Av. Ejercito 441, Santiago, Chile}

\begin{abstract}
Substellar Objects in Nearby Young Clusters -- SONYC -- is a survey program to investigate the frequency and
properties of substellar objects in nearby star-forming regions. We present new spectroscopic follow-up
of candidate members in Chamaeleon-I ($\sim$ 2\,Myr, 160\,pc) and Lupus~3
($\sim$1\,Myr, 200\,pc), identified in our earlier works. We obtained 34 new spectra (1.5 -- 2.4 $\mu$m, R$\sim$600), and identified two probable members in each of the two regions. 
These include 
a new probable brown dwarf in Lupus~3 (NIR spectral type M7.5 and $T_{\mathrm{eff}}=2800$\,K), and
an L3 ($T_{\mathrm{eff}}=2200$\,K) brown dwarf in Cha-I, with the mass below the deuterium-burning limit.
Spectroscopic follow-up of our photometric and proper motion candidates in Lupus~3 is almost complete ($>90\%$), and we conclude
that there are very few new substellar objects left to be found in this region, down to 0.01 -- 0.02\solm~and A$_V\leq 5$. 
The low-mass portion of the mass function in the two clusters can be expressed in the power-law form $dN/dM \propto M^{-\alpha}$, with $\alpha\sim0.7$, in agreement with surveys in other regions. 
In Lupus~3 we observe a possible flattening of the power-law IMF in the substellar regime: this region
seems to produce fewer brown dwarfs relative to other clusters.
The IMF in Cha-I shows a monotonic behavior across the 
deuterium-burning limit,
 consistent with the same power law extending down to 4 -- 9 Jupiter masses. We estimate that objects below the deuterium-burning limit
contribute of the order $5 - 15\%$ to the total number of Cha-I members. 



\end{abstract}
\keywords{stars: formation, low-mass, brown dwarfs, mass function}


\section{Introduction}
\label{intro}

SONYC - short for {\it Substellar Objects in Nearby Young Clusters} - is a comprehensive project aiming
to provide a complete, unbiased census of substellar population down to a few Jupiter masses
in young star forming regions.
The survey is based on extremely deep optical- and near-infrared wide-field imaging, combined
with the Two Micron All Sky Survey (2MASS) and $Spitzer$ photometry catalogs, which are correlated to create 
catalogs of substellar candidates and used to identify targets for extensive spectroscopic follow-up. 
To further facilitate candidate selection, in \citet{muzic14} we also performed a proper motion
analysis of the Lupus~3 star forming region. While the candidate selection based on optical and near-infrared photometry 
helps to avoid biases introduced by the mid-infrared selection 
(only objects with disks), or methane-imaging (only T-dwarfs), it comes with a cost of relatively large candidate lists affected by significant background contamination. 
Spectroscopic follow-up is therefore
a necessary prerequisite to reliably characterize the low mass population in young star forming regions. 

Thanks to our SONYC survey and the efforts of other groups, the substellar Initial Mass Function (IMF) is now well characterized 
down to $\sim$0.01\,\solm. The ratio of the number of stars with respect to brown dwarfs lies between 2 and 6 \citep{scholz13}, and the power-law slope of the mass function is $\alpha$ $\sim$0.6 (dN/dM$\propto$M$^{-\alpha}$; see \citealt{scholz12b, offner13} for a summary). 
It is clear by now that the mass function in young clusters extends below the deuterium-burning (D-burning) limit at $\sim13$M$_{Jup}$, as a handful of early L-type dwarfs have been spectroscopically confirmed in NGC1333 \citep{scholz12b}, $\rho$ Oph \citep{ado12}, $\lambda$~Ori \citep{bayo11}, $\sigma$~Ori \citep{zapatero00, zapatero13}, Orion \citep{ingraham14}, and Cha-I \citep{luhman08}. However, the majority of the current surveys are complete down to $\sim10$M$_{Jup}$, leaving the mass function in the planetary-mass regime still poorly constrained. 
The surveys that extend to lower masses are mostly based on photometry, and still await spectroscopic follow-up.
The SONYC study in NGC~1333 (Scholz et al. 2012b) is unique in this sense, as we obtained spectra of more than 85\% of 
all photometric very-low-mass (VLM) candidates down to $\lesssim 0.005$\solm. We find that the free-floating objects with planetary masses are rare, 20-50 times less numerous than stars 
(for planetary mass objects in range 0.006 - 0.015\solm, and stars below 1\solm). The mass spectrum in NGC~1333 below 0.6\,\solm~is well described by a power-law form 
$dN/dM \propto M^{-\alpha}$, with $\alpha \approx 0.6$, and possibly requires lower values of $\alpha$ in the planetary-mass domain.
In $\sigma$~Ori, \citet{penaramirez12} find the same slope for the mass interval 0.25--0.004 \solm, and 
also suggest a possibly lower $\alpha$ below 0.004\solm. 
While this result still awaits spectroscopic confirmation, three T-dwarf candidates towards $\sigma$~Ori have been 
classified as likely non-members \citep{penaramirez15}, suggesting that the mass function in this cluster might have a
cut-off at $\sim$4 M$_{Jup}$.
In the central part of Upper Scorpius, spectroscopy of L-type candidates \citep{lodieu11} and proper motions of T-type candidates \citep{lodieu13a} favor
a turn-down of the mass function below 10 - 4\,M$_{Jup}$, depending on the age assigned to the cluster, and the models used.
In another photometric survey probing the planetary-mass range in the same region, but over much larger field, 
\citet{lodieu13b} concludes that the mass function (in the log-normal form) is likely decreasing in the planetary-mass regime, although the flat function cannot be discarded at this point. A proper-motion and/or spectroscopic analysis of the candidate members is needed before drawing a firmer conclusion. 

The surveys in star forming regions thus far are possibly in conflict with the claims of the microlensing study by \citet{sumi11}, who suggest that the unbound  or wide-orbit Jupiter-mass objects are almost twice as common as main sequence stars. 
One possible mechanism for the formation of objects of this kind might be ejection of BDs, and/or star forming clumps 
through close encounters in the early stages of cluster formation. However, the typical ejection velocities produced by simulations (e.g. \citealt{bate09, bate12, basu12, reipurth10}) of up to a few km\,s$^{-1}$ might result in a population of BDs distributed at the outskirts of the clusters, but not very far from it. Certainly some objects at the tail of the velocity distribution might escape even further, but this cannot be a substantial number. 
For example, at the end of the cluster simulation in \citet{bate09}, only about 17\% of the low-mass stars and BDs are at the distances beyond the 80\%-mass radius of the cluster.
Therefore, if the microlensing result holds, the objects they probe should have masses below $\sim0.005\,$\solm, and their over-abundances would favor a different formation scenario from the one forming the non-D-burning brown dwarfs in young clusters.


Within SONYC, we have investigated the IMF in the two young star forming regions, Chamaeleon-I (hereafter Cha-I; \citealt{muzic11}), and Lupus~3 \citep{muzic14}.
In this paper we present the second spectroscopic follow-up of the SONYC candidates in these two regions.
Based on this more complete set of follow-up spectra, we can better constrain the properties of the IMF at the low-mass end. In Cha-I specifically, we target 
the candidates expected to have masses below the D-burning limit, in order to derive more definitive conclusions that contribute to the ongoing discussion about the IMF in this mass regime.

In Section \ref{Obs&DR} we give a summary of our previous work in the two regions, describing in detail
the candidate lists and the first spectroscopic follow-up. In Section~\ref{sofintt} we present the new 
spectroscopic follow-up with SofI/NTT. The analysis of the spectra is presented in Section~\ref{dataana}.
The results are discussed in Section~\ref{discuss}. Finally, we summarize the main conclusions in Section~\ref{summary}.

\section{Candidate lists and spectroscopic follow-up}
\label{Obs&DR}

\begin{deluxetable}{lr}
\label{T_samples}
\tabletypesize{\scriptsize}
\tablecaption{Overview of the various samples in Cha-I and Lupus~3 used in this paper}
\tablecolumns{2}
\tablewidth{0pt}
\startdata 
\hline
\\
Sample & No.\tablenotemark{a}\\
\hline
\\
\multicolumn{2}{c}{\bf Chamaeleon-I}\\[2ex]
Candidates selected from $Iz$ diagram & 142/106\\
\hspace{0.3cm} with spectra from VIMOS & 18/18 \\
\hspace{0.3cm} with spectra from SofI & 15/15 \\
\hspace{0.3cm} confirmed VLM objects by VIMOS spectra & 7/7\\
\hspace{0.3cm} confirmed VLM objects by SofI spectra & 2/2\\
\hspace{0.3cm} confirmed VLM objects by other groups\tablenotemark{b} & 9/9 \\
\hspace{0.3cm} rejected as VLM objects by our spectra & 24/24 \\
\hspace{0.3cm} rejected as VLM objects by other groups\tablenotemark{b} & 4/4 \\
\hline
\\
\multicolumn{2}{c}{\bf Lupus 3}\\[2ex]
Candidates selected from $iJ$ diagram\tablenotemark{c} & 372/337 \\
\hspace{0.3cm} with spectra from VIMOS  & 123/112 \\
\hspace{0.3cm} with spectra from SofI & 19/19 \\
\hspace{0.3cm} confirmed young VLM objects by VIMOS spectra & 7/7 \\
\hspace{0.3cm} confirmed young VLM objects by SofI spectra & 2/2 \\
\hspace{0.3cm} confirmed young VLM objects by other groups\tablenotemark{b} & 1/1 \\
\\
Candidates from ``IJ-pm" sample\tablenotemark{d} & 58/53\\
\hspace{0.3cm} with spectra from VIMOS & 32/31 \\
\hspace{0.3cm} with spectra from SofI & 19/19\\
\hspace{0.3cm} confirmed young VLM objects by VIMOS spectra & 7/7\\
\hspace{0.3cm} confirmed young VLM objects by SofI spectra & 2/2 \\
\hspace{0.3cm} confirmed young VLM objects by other groups & 0/0
\enddata
\label{T_samples}
\tablenotetext{a}{total number/above the completeness limit}
\tablenotetext{b}{in addition to the members confirmed/rejected by our spectra}
\tablenotetext{c}{all photometric candidates, including the ``IJ-pm" sample}
\tablenotetext{d}{$iJ$ and proper motion candidates}

\end{deluxetable}



\begin{figure*}
\centering
\resizebox{17cm}{!}{\includegraphics{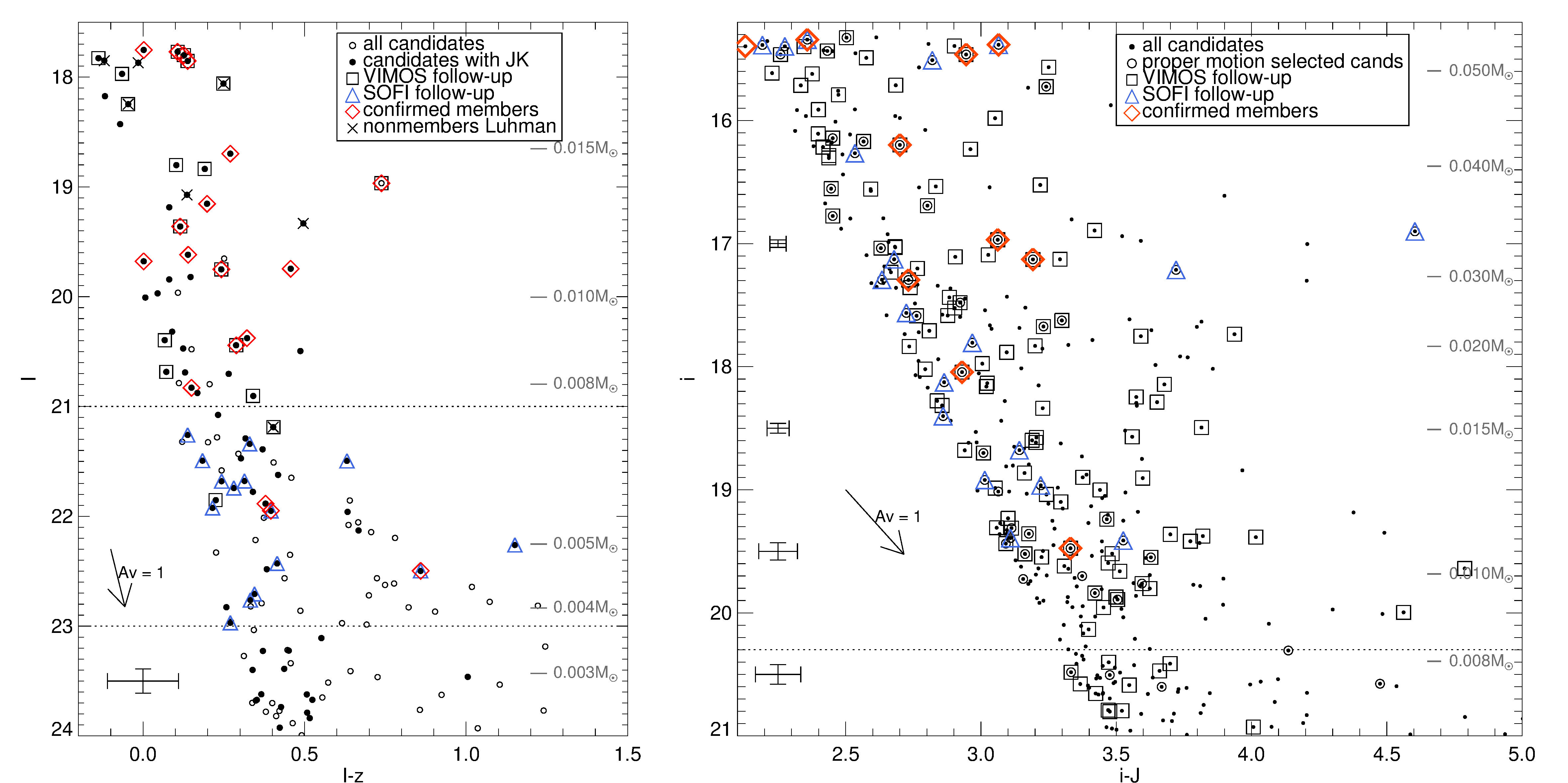}}
\caption{{\bf Left:} Color-magnitude diagram showing candidate members of Cha-I (open circles). The filled circles represent those 
also found in our NIR catalog. The sources observed in two spectroscopic campaigns are shown as black squares (VIMOS/VLT), and blue triangles (SofI/NTT). The confirmed members of Cha-I are marked with red diamonds, and those classified as non-members
in \citet{luhman04} with crosses. The horizontal lines in the left panel mark the upper limit for the SofI follow-up (at $I=21$), and the $I$-band completeness limit ($I=23$). 
{\bf Right:} Color-magnitude diagram showing candidate members in Lupus~3 (dots). Proper motion candidates are additionally marked with open circles. The remaining symbols are the same as in the left panel. $1\sigma$ photometric uncertainties are shown on the left-hand side of the plot. The expected object masses at A$_V=0$ according to the BT-Settl models are shown on the right-hand side of each plot, for
an age of 
2 Myrs in Cha-I, and 1 Myr in Lupus~3.
}
\label{cmd}
\end{figure*}

In this section we outline previous SONYC campaigns in Cha-I \citep{muzic11}, and Lupus~3 \citep{muzic14}, as
the analysis presented in this work is a continuation of our past efforts. We give a summary of the candidate lists for the two regions, 
as well as a detailed description of the second follow-up with SofI/NTT. 
In Table~\ref{T_samples} we provide an overview of the
candidate samples used for the analysis presented in Section~\ref{discuss}. For each entry, we give the total number of objects, and the number of objects that are brighter than the completeness limit in each of the two surveys. The latter is important since we can derive firm statistical conclusions on the shape of the IMF only in the region where our surveys can be considered complete. 
 
\subsection{Chamaeleon-I}

While the Cha-I star forming region spreads over several square degrees on the sky, more than 80\% of the known members 
lie within a $\sim 1\,$deg$^2$ central stripe (see \citealt{luhman07}). The SONYC photometric survey provided deep images
of a part of this area, covering in total $\sim$0.25\,deg$^2$. Our fields contain 38\% of the Cha-I population according to the 
most complete census to date presented in \citet{luhman07}.
Our photometric analysis was based on optical and near-infrared catalogs in $I$, $z$, $J$, and $Ks$ bands\footnote{Throughout this paper, 
$I$ stands for Cousins I, and $i$ for DENIS $i$-band passband. NIR photometry is in the 2MASS system. The $z$-band photometry was 
not absolutely calibrated, and is only used to separate the red from the blue sources in the ($I-z$) color space (see \citealt{muzic11} for details).}. 
A list of candidate members of Cha-I was selected from the ($I-z, I$) color-magnitude diagram (CMD), 
representing a natural extension of the previously known members towards fainter magnitudes. This list contains 142 candidate members,
with $I$-band magnitudes between 17.5 and 24. For Cha-I members within this brightness range, BT-Settl models \citep{allard11} predict masses 
roughly between 0.002 and 0.06\solm, for
A$_V\leq5$\footnote{According to the extinction maps of \citet{cambresy99}, only about 1\% of the Cha-I area has A$_V\geq 5$, while
the entire Lupus~3 area appears to have A$_V<5$. We therefore use this as an upper value of extinction when deriving mass limits from the photometry.}, and an 
age of 1-2 Myr.
The completeness limit of the $I-$band catalog is 23 mag, which is equivalent to $\sim 0.003 - 0.007\, $\solm~for A$_V\leq5$ at 1 Myr, or $\sim 0.004 - 0.009$\solm~at the age of 2 Myrs (median age of Cha-I members when compared to theoretical isochrones; \citealt{luhman07}), and the distance 
of 160 pc.
The completeness limit of the near-infrared catalogs is $J=18.3$ and $Ks=16.7$, which corresponds to masses $\sim 0.004 - 0.007 \, $\solm~for A$_V\leq5$

The first spectroscopic follow-up of the objects in the candidate list was performed using the multi-object spectrograph VIMOS at the VLT. The follow-up included
18 objects from the candidate list, most of which have $I<21$. We confirmed 7 objects as VLM members of Cha-I.
 In the left panel of the Figure~\ref{cmd} we show the CMD for candidate members. 
 For the full CMD we refer the reader to Figure~3 in \citet{muzic11}. Two horizontal lines in the left panel mark the approximate limit of the VIMOS follow-up ($I=21$), and the $I$-band completeness limit ($I=23$).
If indeed members of Cha-I, the objects between these two lines should have masses between 0.003 and 0.015 \solm~at 1Myr or between 0.004 and 0.02~\solm~at 2 Myr (for A$_V\leq5$).
Red diamonds mark spectroscopically confirmed members of Cha-I, from SONYC and \citet{luhman07, luhmanmuench08}, while crosses mark non-members
from \citet{luhman04}.

For the second follow-up with the SofI/NTT, we targeted the sources with expected masses in the planetary-mass regime detected in our NIR images (the filled dots in the left panel
of Figure~\ref{cmd}, located between the two horizontal lines), and not previously observed with VIMOS (squares). The 15 SofI follow-up targets are marked with blue triangles, and listed in 
Table~\ref{T_exp}.






\subsection{Lupus~3}

The SONYC photometric survey in Lupus~3 covers 1.4\,deg$^2$ of the main cluster core, surrounding the two brightest members HR5999/6000. 
Previous surveys by \citet{comeron09}
and \citet{merin08} cover approximately this same area, plus the lower-density part of the complex to the North-East of the main core. This 
part, however, contains only about 10\% of all the candidate members identified in the two aforementioned works.
The original list of photometric candidates in Lupus~3 presented in \citet{muzic14} was selected from the ($i-J$, $i$) CMD, and contained 409 candidates. 
Although in our catalogs we removed the sources too close to the edges of the chips, and overly elongated or blended sources (based on SExtractor output),
some extended sources (galaxies with almost round profiles, unrecognized doubles) and a few artifacts (source sitting right on a bad column, or a spike of 
a nearby bright object) still made it to our candidate catalog. Total of 37 contaminants of this type were found by visual inspection of the images, and have been
removed from the candidate list. The final candidate list then contains 372 sources.
This list can be further narrowed-down to 58, taking into account the proper motions (see \citealt{muzic14} for details). 
As in the previous paper, we refer to this
list as the ``IJ-pm'' candidate list. The photometric completeness limit of our catalog is
$i$=20.3, equivalent to $0.008 - 0.02$ \solm~for $A_V = 0-5$ at a distance of 200 pc and age of 1 Myr, according to the BT-Settl models.
 The magnitude range of the candidate list $I= 15.3 - 21$ is equivalent to masses 0.2 -- 0.007\solm~for $A_V \leq 5$.

In \citet{muzic14}, we presented the first spectroscopic follow-up using VIMOS/VLT, where we took spectra of 123 photometric candidates, out 
of which 32 are from the ``IJ-pm'' list. Spectra confirmed 7 of these objects as young members of Lupus~3, all belonging to the ``IJ-pm'' list.
For the second follow-up, the SofI spectroscopy presented here, the targets were selected from the ``IJ-pm'' list not covered by the VIMOS follow-up, and above the completeness limit. The second follow-up includes 19 candidates, listed in Table~\ref{T_exp}.

The right panel of the Figure~\ref{cmd} shows the CMD containing 372 candidate members of Lupus~3 (dots). The proper motion candidates (``IJ-pm'') are additionally marked with open circles. 
Spectroscopic targets from the two follow-ups (VIMOS and SofI) are marked with black squares and blue triangles, respectively.

\subsection{Spectroscopy with SofI/NTT}
\label{sofintt}

The spectra were taken using SofI (Son of ISAAC; \citealt{moorwood98}) at the ESO's New Technology Telescope (NTT) during four nights
in April 2014, under the program number 093.C-0050(A). 
SofI is equiped with a Hawaii HgCdTe $1024\times1024$ detector, with the pixel size projected onto the sky of $0.288''$.
We used the low resolution red grism (GR Grism Red) covering the wavelength range 
1.5 - 2.5 $\mu$m, and delivering spectral resolution R$\sim 600$ when combined with the $1''$ slit (31 spectra), and 300 with the $2''$ slit (3 spectra). 
The seeing during the SofI observations was varying between 0.7 and $1.5''$ (values from the Differential Image Motion Monitor; DIMM), and the transparency was clear, except
for the first few hours of the night 2014-04-17, when some clouds have been present. The wider slit was used for
the patricularly faint sources ($K \sim 16$), during less favorable seeing conditions. The typical nod throw along the slit was $60''$.
The complete SofI target list, along with the information about the date of observation, on-source integration time, and slit used, is presented
in Table~\ref{T_exp}. 

Data reduction was performed by combining the SOFI pipeline recipes, our IDL routines, and IRAF task $apall$. The data reduction steps include cross-talk, flat field, and distortion corrections. Pairs of frames at different nodding positions were subtracted one from another,
shifted, and combined into a final frame. The spectra were extracted using the IRAF task $apall$, and wavelength calibrated 
 with the help of the Xenon lamp arcs. The wavelength solution obtained by the pipeline has an RMS error of $\sim 20$\AA.
The spectra of telluric standard stars were reduced in the same way as science frames. Telluric standards spectral types range between B5V and B9.5V, and show several prominent hydrogen lines in absorption, which were removed from the spectra by interpolation, prior to division with the black body spectrum at an appropriate effective temperature. This yields the response function of the spectrograph, convolved with the telluric spectrum. 
Finally, we divided the science spectra by the corresponding response function.
The error of the response function is the combination of the errors of
the measured telluric spectra and the errors in the effective temperature assigned to certain spectral types. 
The latter is 
caused by 
assumption that the instrinsic spectrum of the calibrating star is well 
described by a black body of a certain temperature. 
The error of the measured telluric spectra is supplied by $apall$, through the $extras$ keyword. It is of the order, or below 1\% across most of the spectrum, except in the region between H and K-bands where the atmosphere absorbs most of the light. 
For the B-type telluric standards, we adopted the temperature scale from \citet{boehm81}, whose $\pm 5\%$ error on the effective temperature
translates to approximately $\pm 1\%$ error in the response curve.

\begin{deluxetable*}{rlllclllll}
\tabletypesize{\scriptsize}
\tablecaption{Photometric candidates included in the spectroscopic follow-up with SofI}
\tablewidth{0pt}
\tablehead{\colhead{$\#$} &
	   \colhead{$\alpha$(J2000)} & 
	   \colhead{$\delta$(J2000)} & 
	   \colhead{Date} & 
	   \colhead{Exp. time} &
	   \colhead{Slit}	& 
	   \colhead{$I$\tablenotemark{a}} & 
	   \colhead{$J$} &
	   \colhead{$K$} & 
 	   \colhead{comments}}
\tablecolumns{10}
\startdata 
\multicolumn{10}{c}{\bf Chamaeleon-I}\\[2ex]
\hline
\\
1 & 11 05 21.36 & -77 46 33.0 & 2014-04-19 & $14 \times 180\,$s & $1''$ & 21.92  & 17.37 &      14.65 \\
2 & 11 06 18.62 & -77 35 17.4 & 2014-04-18 & $12 \times 180\,$s & $1''$ & 22.43   & 16.70 &     13.22  \\ 
3 & 11 06 28.04 & -77 35 54.7 & 2014-04-13 & $14 \times 60\,$s  & $1''$ & 21.68  & 16.50 &      13.31\\ 
4 & 11 08 20.20 & -77 45 48.6 & 2014-04-13 & $10 \times 200\,$s & $1''$ & 21.50  & 18.09 &      15.86 \\ 
5 & 11 08 23.59 & -77 31 29.5 & 2014-04-14 & $52 \times 100\,$s & $2''$ & 22.26   & 18.65 &      16.49\\
6 & 11 08 30.31 & -77 31 38.6 & 2014-04-14 & $52 \times 100\,$s & $2''$ & 22.50 & 17.92 & 16.05 &  confirmed member		\\ 
7 & 11 08 30.77 & -77 45 50.5 & 2014-04-13 & $10 \times 200\,$s & $1''$ & 21.34  & 17.77 &      15.59 \\ 
8 & 11 08 33.92 & -77 46 33.4 & 2014-04-13 & $14 \times 220\,$s & $1''$ & 21.49  & 18.54 &      16.68\\ 
9 & 11 08 50.38 & -77 45 53.0 & 2014-04-13 & $14 \times 220\,$s & $1''$ & 21.26  & 17.90 &     16.01\\ 
10 & 11 09 11.62 & -77 19 26.1 & 2014-04-18 & $10 \times 180\,$s & $1''$ & 21.74  & 16.56 &      13.20\\
11 & 11 09 28.57 & -76 33 28.1 & 2014-04-14 & $18 \times 90\,$s & $1''$ & 21.95 & 16.74 & 11.93 &  confirmed member\\ 
12 & 11 09 32.26 & -76 34 39.3 & 2014-04-19 &  $20 \times 90\,$s & $2''$ & 22.76   & 18.52 &      15.74\\
13 & 11 09 35.44 & -77 17 27.1 & 2014-04-18 & $8 \times 180\,$s & $1''$ & 21.68  & 16.73 &      13.60 \\ 
14 & 11 09 37.38 & -76 41 05.7 & 2014-04-17 & $16 \times 250\,$s  & $1''$ & 22.71   & 17.92 &      15.18\\
15 & 11 09 51.82 & -76 42 18.2 & 2014-04-19 & $12 \times 180\,$s & $1''$ & 22.97   & 17.93 &      14.57\\
\hline
\\
\multicolumn{10}{c}{\bf Lupus 3}\\[2ex]
\hline

\\
16 & 16 08 03.96 & -39 10  28.9 & 2014-04-18 & $10 \times 150\,$s & $1''$ & 19.39 & 16.28  & 14.21 \\ 
17 & 16 08 25.87 & -39 10  08.0 & 2014-04-17 & $8 \times 120\,$s  & $1''$ & 18.13 & 15.26  & 13.03 \\ 
18 & 16 08 30.04 & -39 10  01.4 & 2014-04-17 & $8 \times 120\,$s & $1''$ & 15.51 & 12.69  & 10.82\\
19 & 16 08 33.04 & -38 52 22.7  & 2014-04-18 &  $6 \times 180\,$s & $1''$ & 15.34 & 12.98 & 11.74 &  confirmed member\\ 
20 & 16 09 42.66 & -39 03  20.7 & 2014-04-18 & $10 \times 180\,$s & $1''$ & 18.68 & 15.54  & 13.29 \\ 
21 & 16 09 43.74 & -39 07  11.9 & 2014-04-17 & $12 \times 120\,$s & $1''$ & 18.96 & 15.74  & 13.23 \\ 
22 & 16 09 47.20 & -39 08  42.4 & 2014-04-19 &  $6 \times 180\,$s & $1''$ & 18.92 & 15.91  & 13.82\\ 
23 & 16 09 52.68 & -39 06  40.4 & 2014-04-17 & $8 \times 120\,$s & $1''$ & 17.13 & 14.45  & 12.52 \\ 
24 & 16 10 01.33 & -39 06 45.1  & 2014-04-17	& $8 \times 60\,$s & $1''$ & 15.38 & 12.32 & 10.52 & confirmed member		    \\ 
25 & 16 10 05.97 & -39 08  06.2 & 2014-04-18 & $10 \times 180\,$s & $1''$ & 18.40 & 15.54  & 13.29\\ 
26 & 16 10 06.01 & -39 04  23.7 & 2014-04-17 & $6 \times 60\,$s & $1''$ & 16.90 & 12.30  &  9.59 & giant? \\ 
27 & 16 10 11.71 & -39 09  19.8 & 2014-04-17 & $8 \times 180\,$s  & $1''$  & 17.81 & 14.84  & 12.96 \\ 
28 & 16 10 13.74 & -39 02  34.8 & 2014-04-17 & $10 \times 120\,$s & $1''$ & 17.21 & 13.49  & 10.86 & giant? \\
29 & 16 10 14.59 & -39 02  31.4 & 2014-04-17 & $10 \times 120\,$s & $1''$ & 19.41 & 15.89  & 13.58 \\
30 & 16 10 17.86 & -39 03  47.1 & 2014-04-17 & $6 \times 120\,$s & $1''$ & 16.26 & 13.73  & 11.93\\ 
31 & 16 10 30.82 & -39 04  04.4 & 2014-04-18 & $6 \times 180\,$s  & $1''$ & 17.29 & 14.66  & 12.81\\ 
32 & 16 10 33.59 & -39 08  21.4 & 2014-04-17 & $6 \times 180\,$s & $1''$ & 17.56 & 14.84  & 12.54 \\
33 & 16 11 46.57 & -39 06  04.7 & 2014-04-17 & $8 \times 90\,$s & $1''$ & 15.39 & 13.19  & 11.23 & member? \\ 
34 & 16 12 04.95 & -39 06  25.6 & 2014-04-17 & $6 \times 120\,$s & $1''$ & 15.40 & 13.12  & 11.51 & giant or member 
\enddata
\label{T_exp}
\tablenotetext{a}{Cousins I for sources in Chamaeleon, and DENIS $i$ for those in Lupus}
\end{deluxetable*}

\section{Data analysis}
\label{dataana}

The analysis of the spectra presented in this section consists of: (1) visual inspection of the spectra in search for 
the features revealing the youth, (2) fitting of the model spectra to the observed ones to determine the effective temperature and extinction, and 
(3) determination of spectral types.   

\subsection{Membership assessment}
\label{S_membership}

\begin{figure}
\centering
\resizebox{9cm}{!}{\includegraphics{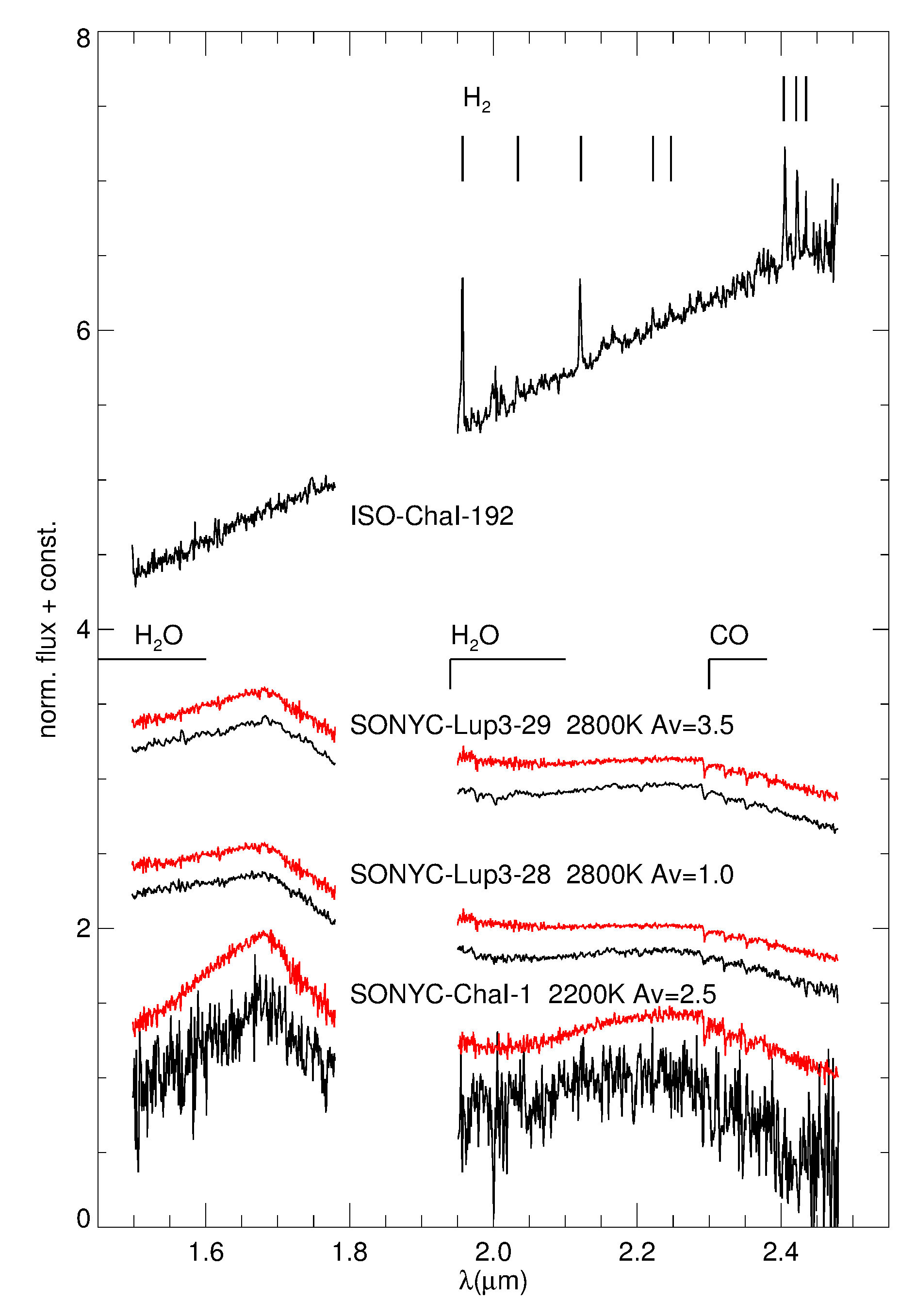}}
\caption{Spectra of the confirmed members of Cha-I and Lupus~3 (black), along with the BT-Settl best-fit models shown in red. All the spectra
are normalized at 1.6$\mu$m, and shifted by arbitrary offsets for clarity. The object spectra are as observed, and the models have been 
artificially reddened. The region strongly affected by telluric absorption (1.8 -- 1.95\,$\mu$m) is not shown for clarity. The prominent
molecular absorption bands observed in ultracool objects are marked. The lines marked at the top of the plot are the H$_2$ ro-vibrational transitions.
}
\label{F_conf}
\end{figure}

\begin{figure}
\centering
\resizebox{9cm}{!}{\includegraphics{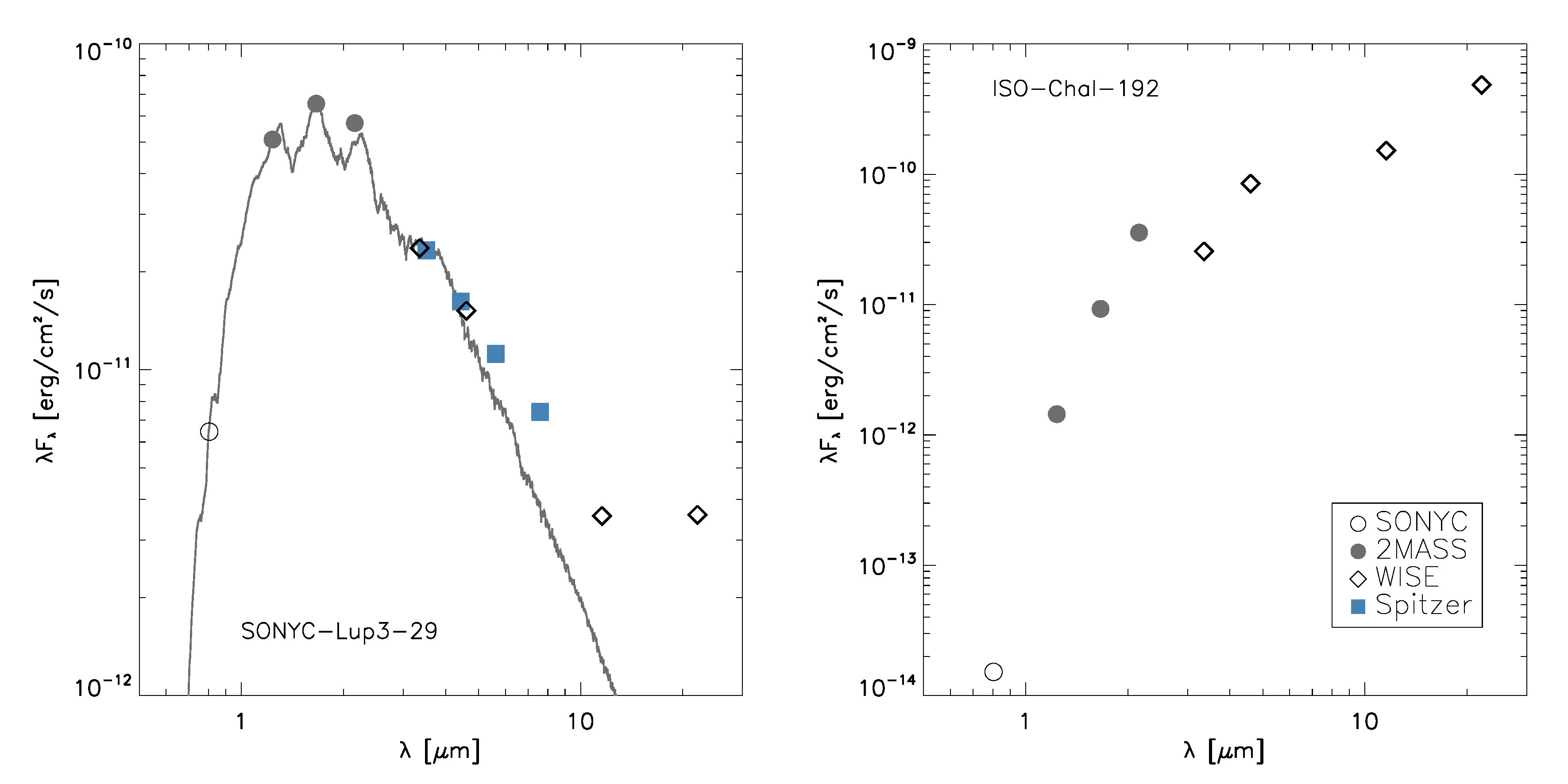}}
\caption{Spectral energy distribution for two objects from our follow-up showing infrared excess. The solid line
in the left panel is a BT-Settl atmospheric model for a low-gravity object with $T_{\mathrm{eff}}$=2800\,K.}
\label{SEDs}
\end{figure}

\begin{figure*}
\centering
\resizebox{13cm}{!}{\includegraphics{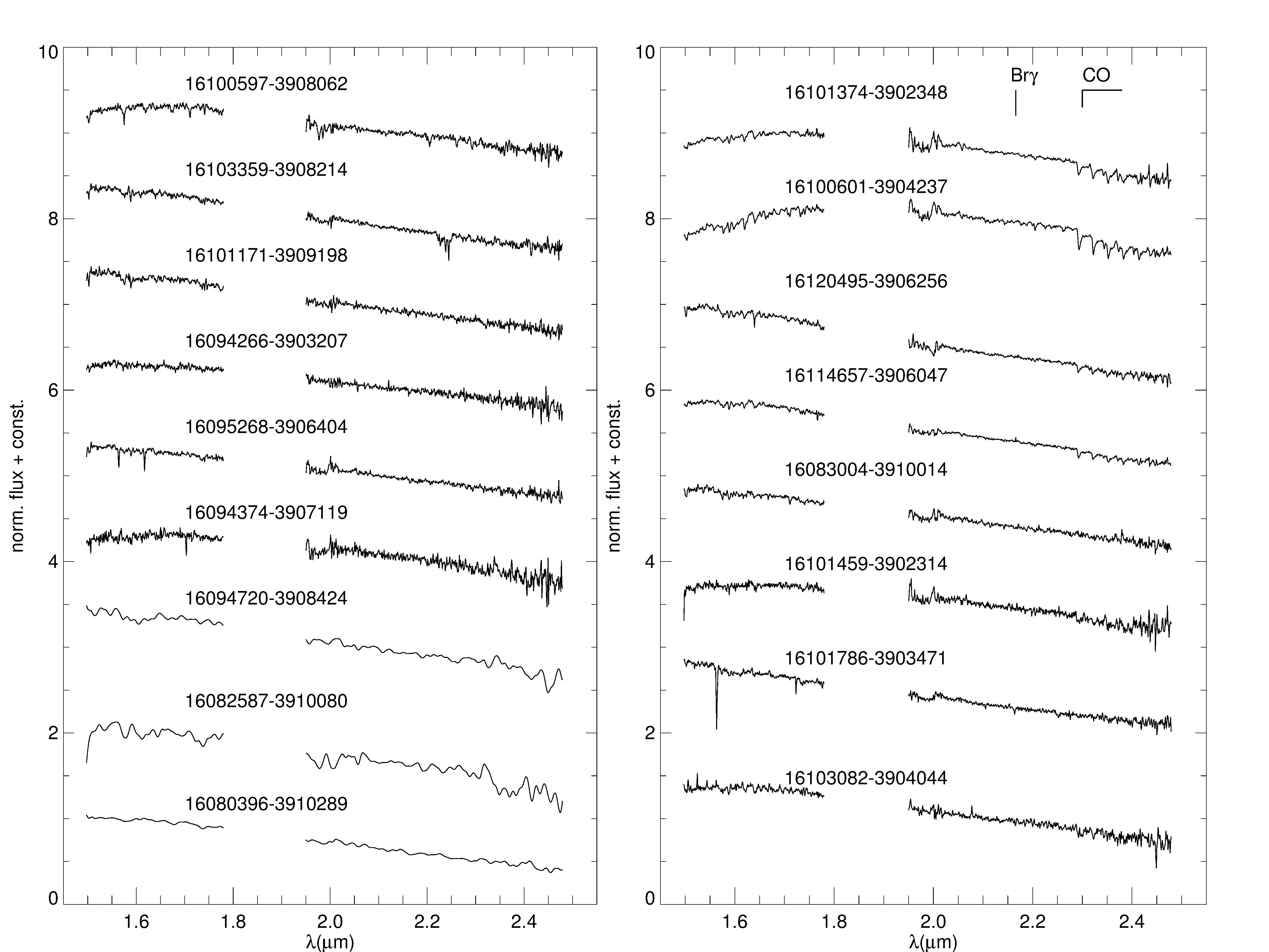}}
\caption{Observed spectra of candidate objects in the direction of Lupus~3, for which the membership cannot
be determined with certainty. While it is clear that none of these objects is a young substellar member,  
it cannot be excluded that some of these still could be stellar members with spectral types earlier than $\sim\,$M5.
  All the spectra
are normalized at 1.6$\mu$m, and shifted by arbitrary offsets for clarity. Three spectra shown at the bottom of the left panel have lower
S/N and were smoothed to better appreciate the overall broad shape of the spectra. The region strongly affected by telluric absorption (1.8 -- 1.95\,$\mu$m) is not shown for clarity. 
}
\label{rest_lup}
\end{figure*}

\begin{figure*}
\centering
\resizebox{13cm}{!}{\includegraphics{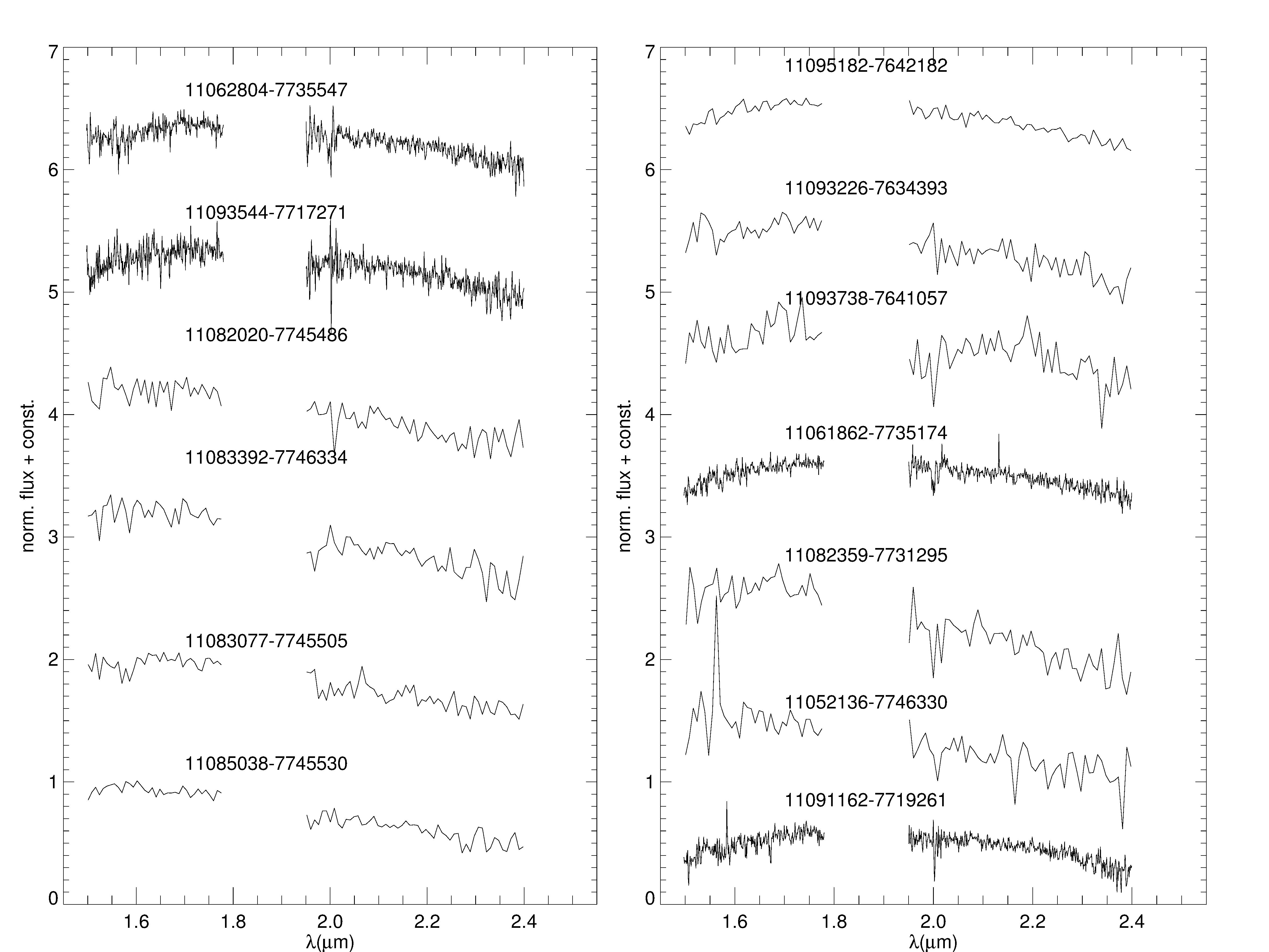}}
\caption{Observed spectra of candidate objects in the direction of Cha-I, for which the membership cannot
be determined with certainty. While it is clear that none of these objects is a young substellar member,  
it cannot be excluded that some of these still could be stellar members with spectral types earlier than $\sim\,$M5. All the spectra
are normalized at 1.6$\mu$m, and shifted by arbitrary offsets for clarity. With four exceptions, the spectra were smoothed to a lower resolution to suppress the noise. The region strongly affected by telluric absorption (1.8 -- 1.95\,$\mu$m) is not shown for clarity.
}
\label{rest_chai}
\end{figure*}

The initial assessment of the spectra is done by visual inspection. Spectra of young objects later than $\sim\,$M5 show a characteristic triangular peak in the H-band, 
caused by water absorption \citep{lucas01,cushing05}. The depth of this feature strongly depends on effective temperature, while the shape is affected by the gravity. The peak appears
triangular for young, low- to intermediate-gravity\footnote{$\lesssim 300$\,Myr according to \citet{allers13}.} objects, and round at higher gravities. The K-band spectra of late type objects show prominent CO absorption bands. CO bands can facilitate 
spectral classification, since K and M-type giants show deeper CO absorption than dwarfs (see e.g. \citealt{rayner09}). We also look for the emission lines, which may indicate accretion. Only three objects from the SofI spectroscopic sample (1 in Cha-I, 2 in Lupus~3) clearly show a triangular H-band peak. They are shown 
in Table~\ref{T_conf}, and Figure~\ref{F_conf} (black), along with the corresponding best fit models (red; see the next section). SONYC-Lup3-29
 is a new, probably sub-stellar member of Lupus~3. The membership is further confirmed by the mid-infrared excess, as demonstrated in Figure~\ref{SEDs}. 

In Figure~\ref{F_conf}, we also show the spectrum of ISO-ChaI-192, whose steep spectrum with prominent H$_2$ emission lines indicates a YSO nature \footnote{The 
clearly detected H$_2$ lines are 1-0 S(2) at 2.034$\mu$m, 1-0 S(1) at 2.122$\mu$m, 1-0 S(3) at 1.957$\mu$m, 1-0 S(0) at 2.222$\mu$m, 2-1 S(1) at 2.247$\mu$m, 
1-0 Q(1) at 2.404$\mu$m, 1-0 Q(3) at 2.421$\mu$m, and 1-0 Q(4) at 2.435$\mu$m.}.
Indeed, this object is a known 
class I protostar and an FU Ori candidate \citep{gramajo14, cambresy98, persi99}, associated with a CO outflow \citep{mattila89, persi07}. 
\citet{gomez03} derive spectral type of M3.5-M6.5 from the near infrared low-resolution spectra.
Two estimates for the effective temperature are found in the literature, 3600$\,$K by \citet{persi07}, and 5000$\,$K by \citet{gramajo14}.
The typical class-I SED of ISO-ChaI-192 is shown in Figure~\ref{SEDs}.

The four probable members and their properties derived in the next sections are listed in Table~\ref{T_conf}.
The remaining objects show flat spectra with few features, and cannot be easily classified at the resolution and signal-to-noise of our spectra. 
They are shown in Figures~\ref{rest_lup} and \ref{rest_chai}, for
Lupus~3, and Cha-I, respectively, and listed in Table~\ref{T_exp} (all objects in this table except those with the comment ``confirmed member"). 
The fact that their spectra do not show the water absorption in the H-band indicative of cool spectral types
means that they must have spectral type earlier than $\sim$M5, i.e. they are certainly not substellar. 
The lack of CO bands at 2.3$\,\mu$m in most of these spectra is another signpost of warm atmospheres.
Only the top-most four spectra of the right panel in Figure~\ref{rest_lup} clearly show CO absorption bands. Comparison with the 
giant and dwarfs spectra from the IRTF spectral library \citep{rayner09}, reveal that they are most probably K-type giant stars, although in the case of 
$\#33$ (16114657-3906047) and $\#34$ (16120495-3906256)  we cannot with certainty exclude that they are young, since the CO bands are not as deep as in the other two objects. Additionally, 
$\#33$ exhibits Br$\gamma$ in emission, i.e. this object might be a stellar member of Lupus~3. 
For the remaining objects listed in Table~\ref{T_exp} without any comment, the low resolution and S/N of our spectra prevents
a straightforward classification. 
\subsection{Model fitting}
\label{S_modelfit}
To determine the effective temperature and extinction of the three objects showing the prominent H-band peak, 
we fitted the spectra with the BT-Settl models \citep{allard11}. The procedure is identical to the one
applied in other papers of the SONYC series using NIR spectroscopy (for a detailed description of the procedure, see \citealt{scholz12a}). 
The best fit solution is searched on a grid of $T_{\mathrm{eff}}$ varied between 2000 and 4000$\,$K in steps of 100$\,$K, and A$_V$ within $\pm 5\,$mag of 
the value derived from photometry, by assuming the intrinsic $(J -K)_0=1$, and extinction law from \citet{cardelli89}\footnote{As argued in 
\citet{scholz09}, the value $(J -K)_0 = 1$ is appropriate for
objects of the M spectral type. This is in agreement with values for 5-30 Myr old M-dwarfs by \citet{pecaut13}.}.
 The log$g$ is kept at a single value of 3.5, suitable for young objects that are still contracting.
For the two objects in Lupus~3 no, or a very small adjustment to the photometric value of extinction ($\Delta A_V<1$) 
is needed to produce a satisfying fit. For SONYC-ChaI-1, on the other hand, the value about $2-2.5\,$mag lower provides a much better solution. 
For the uncertainties of the fitting procedure, we adopt the conservative values of $\pm200$\,K for the
$T_{\mathrm{eff}}$, and $\pm 1$\,mag for the A$_V$, same as in our earlier works.
SONYC-Lup3-28 was previously confirmed as a probable 
substellar member of Lupus~3 by \citet{comeron13}, who derived the $T_{\mathrm{eff}} = 2800$\,K, in agreement with the value derived here.
SONYC-ChaI-1 was previously reported by \citet{luhman07}, and classified as a probable member of Cha-I, with spectral type $\geq$M9, and $T_{\mathrm{eff}}\leq$2400$\,$K, in agreement with 
our results.

\subsection{Spectral types}

\begin{figure}
\resizebox{8.5cm}{!}{\includegraphics{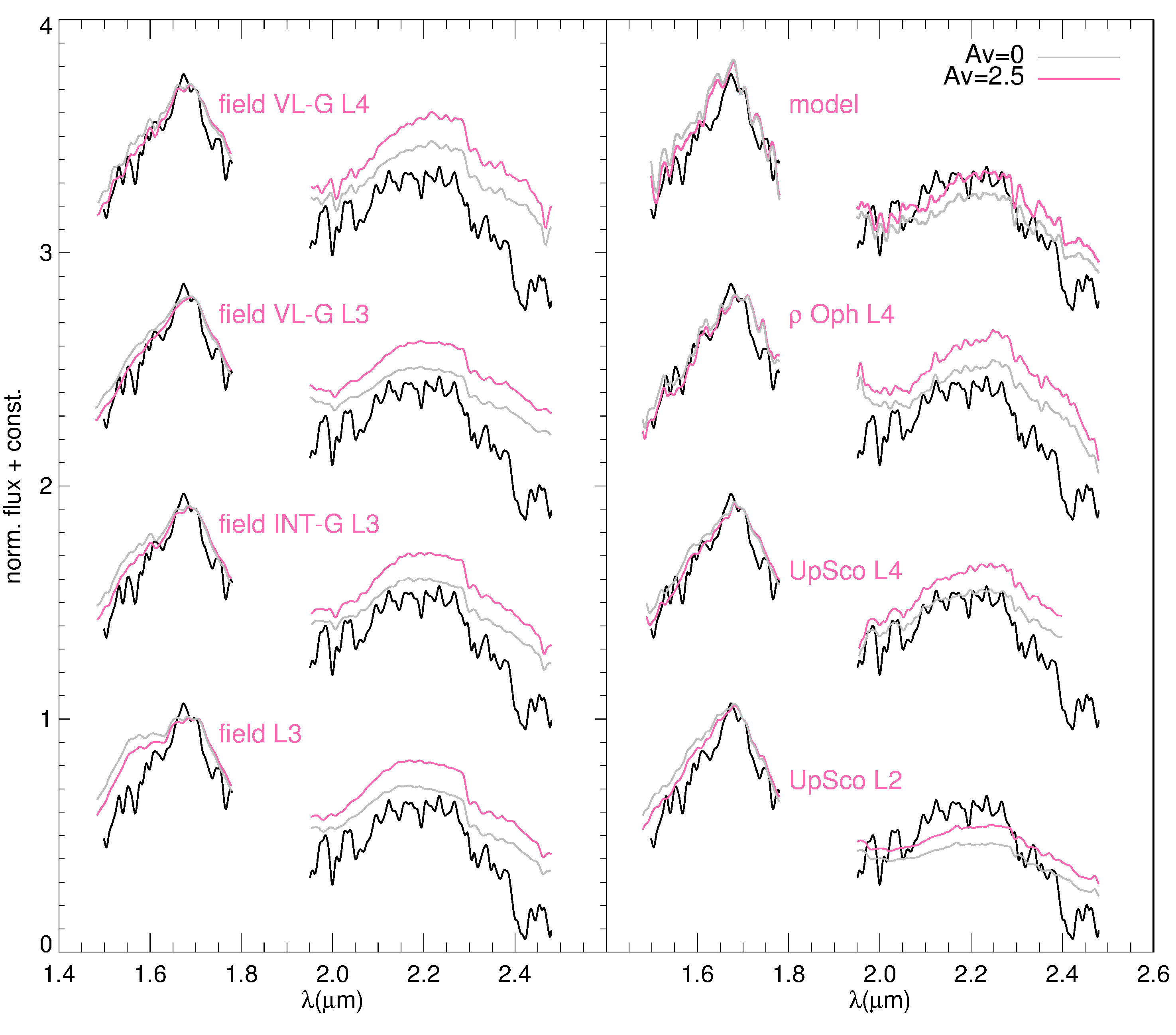}}
\centering
\caption{Comparison of the spectrum of SONYC-ChaI-1 (black) with the spectra of field ultracool dwarfs from \citet[][left panel]{allers13}, and 
young L-type members of star forming regions (right panel). Spectra of the comparison objects are shown with no reddening (grey), and with the extinction of A$_V =2.5$ (the value estimated for 
SONYC-ChaI-1 from the model fitting; magenta) applied to it.
The field objects are, from bottom to top, 2MASSJ0407+1546 (L3 field gravity),
 2MASSJ1726+1538 (L3 intermediate gravity), 2MASSJ1615+4953 (L3 very-low gravity), and 2MASSJ15515237+0941148 (L4 very-low gravity).
The young objects include two members of Upper Sco, UScoJ160603-221930 (L2; \citealt{lodieu08}) and 1RXSJ1609-2105B (L4; \citealt{lafreniere10}), and CFHTWIR-Oph33, a member of $\rho\,$Oph (L4; \citealt{ado12}). For a reference, we also show the best-fit BT-Settl low-gravity model (2200 K).
All spectra are smoothed to the same resolution, and normalized at 1.7$\mu$m.
}
\label{Ls}
\end{figure}

We derived spectral types for the three late type members confirmed in this work, using the following spectral indices: 
(1) H-peak index (HPI, \citealt{scholz12a}), (2) H$_2$O index \citep{allers07}, and (3) Q-index \citep{wilking99}. 
The three indices give consistent NIR spectral types for the two M-dwarfs found in Lupus~3. 
For SONYC-Lup3-28 we get M8 (HPI and Q-index), 
and M7 (H$_2$O), in agreement with M7.5 derived by \citet{comeron13}.
 For SONYC-Lup3-29 we get M8 (HPI), M6 (H$_2$O), and M7 (Q). For SONYC-ChaI-1, the indices result in L2 (H$_2$O), 
and L4 (HPI and Q-index). While all the three indices are defined for young BDs earlier than L0, \citet{scholz12a} note that HPI might
hold also at later spectral types. \citet{allers07} derived two SpT-H$_2$O index relations, one for young BDs in the M5 -- L0 range, and other for the field dwarfs in
M5 -- L5 SpT range. They note that the two relations appear very similar, and thus the SpT-index relation can also be used for the young
BDs later than L0.  
\citet{luhman07} derive spectral type $\geq$M9 for SONYC-ChaI-1.

Prior to the calculation of the spectral types, the spectra had to be corrected for extinction. 
To determine the influence of extinction on the resulting spectral types, we varied the A$_V$ derived in the previous section by $\pm 1$\,mag. 
For HPI, we obtain the uncertainty of $\pm 0.3$ subtypes for SONYC-Lup3-28 and -29, and $\pm 0.5$ for the noisier SONYC-ChaI-1. For
H$_2$O index, we get $\pm 0.3$ for all three objects, while the Q-index does not depend on the extinction. 
Additionally, the uncertainties for the
spectral types derived from the three indices from their respective defining papers are $\pm0.4$ for HPI, 
$\pm 1.0$ for the H$_2$O-index, and $\pm 1.5$ for the Q-index.
These uncertainties have been added in quadrature to the uncertainties from the extinction, and taken into
account for the calculation of the final spectral types, which were rounded-up to the nearest half-spectral type.

In Figure~\ref{Ls}, we show a comparison of the SONYC-ChaI-1 spectrum with the spectra of field ultracool dwarfs from \citet{allers13}, and with the young L-type members in Upper Sco and $\rho\,$Oph. \citet{allers13} presented a spectroscopic study of field ultracool dwarfs having spectroscopic and/or kinematic evidence of youth ($\approx 10 -300$ Myr). The objects are divided in three gravity classes based on the shape and strength of various features in their spectra. Three L3 objects exhibiting high-, intermediate-, and very-low gravity features are shown
in the left panel of Figure~\ref{Ls}. At the top of the same panel we also show a very-low gravity L4 object to facilitate the comparison with the young members of star forming regions, shown in the right panel. There are no suitable intermediate-gravity L4 standards presented in \citet{allers13}. 
From the shape of the H-band peak, it is evident that SONYC-ChaI-1 cannot be a normal field dwarf, but it is 
difficult to judge whether an intermediate or a low-gravity atmosphere provides a better fit. 
 Overall, the spectral features match well both L3 or L4 low-gravity objects, but the slopes between H and K bands are different, in the sense that
our object appears bluer than any of the field templates, even without extinction.  

To date, there is only a limited sample of very young L-type members of star forming regions with 
good quality spectra available for comparison, especially at spectral types later than L1. In the right panel of Figure~\ref{Ls}, we show two members of the Upper Scorpius association, UScoJ160603-221930 (L2; \citealt{lodieu08}) and 1RXSJ1609-2105B (L4; \citealt{lafreniere10}), along with CFHTWIR-Oph33, an L4 member of $\rho$ Oph \citep{ado12}.
The H-band portion of the SONYC-ChaI-1 spectrum matches well the young L4 templates, but in the K-band it again appears to be more consistent with 
A$_V=0$, rather than  with A$_V=2.5$, the value estimated from the spectral model fitting. For a comparison, on top of the right panel we show 
the best-fit BT-Settl model, which clearly prefers the higher value of the extinction.

We note that detailed spectral type comparisons at young ages can be complicated by possible excess emission from disks or accretion. 
As demonstrated by \citet{dawson14}, these characteristics alter the spectra of Class II objects, which then show different normalized levels between individual NIR bands. The effect is also present in Class III object spectra, to a somewhat lesser extent. 

For now, we conclude that the spectrum of SONYC-ChaI-1 is consistent with a spectral type L3-L4, A$_V \lesssim 3$, and 
a low-gravity atmosphere. 
More high quality spectra of young L-type objects are needed in order to establish a proper NIR spectral type sequence, and to create a gravity sequence at each spectral type, in comparison to 
the field ultracool low-gravity dwarfs from \citet{allers13}.

\begin{deluxetable*}{cllccccclll}
\tabletypesize{\scriptsize}
\tablecaption{Confirmed members of Cha-I and Lupus~3, from the SofI follow-up.}
\tablewidth{0pt}
\tablehead{\colhead{No.\tablenotemark{a}} &
	   \colhead{name} & 
	   \colhead{$\alpha$(J2000)} & 
	   \colhead{$\delta$(J2000)} & 
	   \colhead{SpT\tablenotemark{b}} & 
	   \colhead{$T_{\mathrm{eff}}$/K} &
	   \colhead{A$_{V}$/mag} &
	   \colhead{mass/\solm\tablenotemark{c}} &
	   \colhead{log(L/\soll)} &
 	   \colhead{comments}} 
\tablecolumns{10}
\startdata 
6 & SONYC-ChaI-1  & 11 08 30.31 & -77 31 38.6  & L$3\pm0.5$ & 2200 & 2.5 & 0.009-0.012 & $-3.44 \pm 0.10$ & $\geq$M9, $\leq$2400K\tablenotemark{d} \\ 
11 & ISO-ChaI-192 & 11 09 28.57 & -76 33 28.1  & \nodata   & \nodata & \nodata  & \nodata  & $-0.11 \pm 0.20$&  \\ 
19 & SONYC-Lup3-28 & 16 08 33.04 & -38 52 22.7 &  M$8\pm0.5$ & 2800 & 1.0 & 0.05-0.06 & $-1.37 \pm 0.11$& M7.5, 2800$\,$K\tablenotemark{e} \\ 
24 & SONYC-Lup3-29 & 16 10 01.33 & -39 06 45.1 & M$7.5\pm0.5$ & 2800 & 3.5 & 0.05-0.06 & $-0.81 \pm 0.11$ & IR excess
\enddata
\label{T_conf}
\tablenotetext{a}{Same as in Table~\ref{T_exp}.}
\tablenotetext{b}{NIR spectral type}
\tablenotetext{c}{estimate based on $T_{\mathrm{eff}}$ and BT-Settl models at 1\,Myr (Lupus3), and 2\,Myr (Cha-I)}
\tablenotetext{d}{\citet{luhman07}}
\tablenotetext{e}{\citet{comeron13}}
\end{deluxetable*}

\subsection{Hertzsprung-Russell Diagram}
\label{S_HRD}

\begin{figure*}
\centering
\resizebox{15cm}{!}{\includegraphics{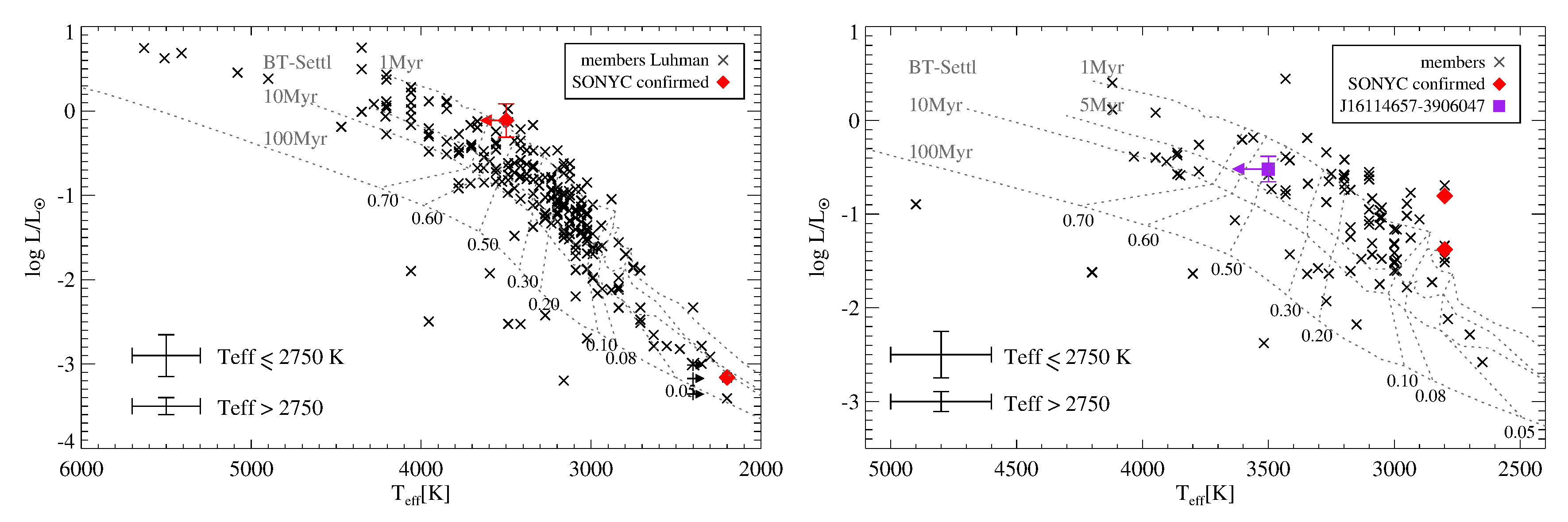}}
\caption{H-R diagram for Cha-I (left) and Lupus$\,$3 (right). The crosses mark the members of ChaI from
\citet{luhman07,luhman08,luhmanmuench08} and Lupus~3 from the census table in \citet{muzic14}.
Red diamonds mark the members confirmed in this work, and the purple square marks the candidate member of Lupus~3
16114657-3906047. The arrows signal the lower or upper limits on $T_{\mathrm{eff}}$.
The dashed lines show BT-Settl theoretical tracks for ages of 1, 5, 10, and 100 Myrs. 
}
\label{HR}
\end{figure*}

In Figure~\ref{HR}, we show HR diagrams for Cha-I (left panel), and Lupus~3 (right panel). 
The luminosities were calculated from the $J$-band magnitudes corrected for extinction, distance modulus, and bolometric correction.
We adopt a distance of 200 pc for Lupus~3, and 160 pc for Cha-I. Extinction was calculated from the $J-K$ colors, 
assuming the intrinsic colors according to spectral type of each member, as listed in \citet{pecaut13} for young (5-30 Myr) dwarf stars.
We use the extinction law from \citet{cardelli89}, with R$_V$ = 4.
The bolometric correction in the J band (BC$_J$), for the objects with $T_{\mathrm{eff}} > 2750$\,K, is calculated from the polynomial
relation between BC$_J$ and $T_{\mathrm{eff}}$ derived by \citet{pecaut13}. For the objects with $T_{\mathrm{eff}} \leq 2750$\,K, where this
relation is not valid, we determine
the BC$_J$ as a function of $J-K_S$ color, as derived in \citet{schmidt14}. The former relation is valid for young (5-30 Myr) dwarfs with
spectral type earlier than M6, while the
latter was derived for field dwarfs with spectral types M7-L8.

The HR diagram for Lupus~3 is an updated version of the one presented in \citet{muzic14}, where we added the two objects confirmed in
this work (red diamonds), and the candidate member J16114657-3906047 (purple square). For the latter we can only infer a lower limit on
the $T_{\mathrm{eff}}$. Since we do not know the exact spectral type of this object, we adopt the intrinsic $J-K$ color 0.6-0.9 (used to
calculate the A$_V$), and J-band bolometric correction $1.5\pm0.2\,$mag, both suitable for young K-type stars \citep{pecaut13}. To construct the 
Cha-I HR diagram, we use the photometry, $T_{\mathrm{eff}}$, and extinction from the census by \citet{luhman07}, add the objects from \citet{luhman08,luhmanmuench08}, 
and the two sources confirmed in this work (red diamonds).
The typical error-bars are shown in the lower left corner of the two plots. 
 $T_{\mathrm{eff}}$ estimates for Lupus~3 come from various works, and we take $\pm 200$K as a representative uncertainty 
(see \citealt{muzic14} for details). 
The $T_{\mathrm{eff}}$ of the Cha-I members marked with crosses comes from the Luhman census, 
who converts spectral types to $T_{\mathrm{eff}}$ using the scale from \citet{kenyon95} for the stars with spectral types earlier than M, 
and the one from \citet{luhman03b} for the M-type. The latter scale was designed to be compatible with the evolutionary models of \citet{baraffe98}. 
Therefore, in addition to the uncertainties in spectral types, these temperature estimates
are subject to a systematic uncertainty in the temperature scale, which is difficult to predict, since it depends on various details of the models, and is probably at least $\pm 100$\,K \citep{luhmanmuench08}. We therefore decide to adopt a single value of $\pm200$\,K for all objects in the plot.
Different uncertainties in luminosity for objects hotter, or cooler than 2750\,K in Figure~\ref{HR}, result from the uncertainties in BC$_J$
that were adopted from two different works \citep{pecaut13,schmidt14}.
The two sources with uncertain spectral type 
have slightly larger log$L$ uncertainty, as a consequence of a range of extinctions and bolometric magnitudes. 

In both plots, there are sources appearing well below the main sequence formed by the cluster members. In case of Cha-I, for 7 out of the 8 sources with $T_{\mathrm{eff}}>3000\,$K that are located below the 100 Myr isochrone, \citet{luhman04,luhman07} argues in favor of these sources being genuine members of Cha-I, based on mid-IR excess and/or strong emission lines. Only one of the 8 sources, OTS 32, has an uncertain membership. 
In Lupus~3, 3 of the underluminous objects also show presence of a disk and strong emission lines, while one of them
shows strong H$_{\alpha}$ emission, but no mid-IR excess \citep{muzic14, comeron03, fernandez05}.  
As discussed in \citet{comeron03}, the underluminosity of the young members of star forming regions can be explained by the object being
seen only through scattered light (edge-on disk, or embedded Class I sources), or by the accretion-modified evolution, similar to what is described
in \citet{hartmann98} and \citet{baraffe09}.
The examples of the detail modeling of young underluminous objects with edge-on disks can be found in e.g. \citet{scholz10, huelamo10, petr-gotzens10}. 

In Cha-I, the sequence of the objects hotter than $\sim2700\,$K is following the isochrones, and is mostly confined between those marking 1 and 10 Myr. 
Most of the coolest objects, on the other hand, fall between 10 and 100 Myr, i.e. they 
appear systematically underluminous relative to theoretical isochrones. 
One possible culprit are the bolometric corrections used to calculate the luminosities. To calculate BC$_J$ for these objects, we use the relation from \citet{schmidt14}, which was derived for the field dwarfs. 
As summarized previously in \citet{luhman12}, the bolometric corrections for young L and T dwarfs
are probably not the same as those for standard field dwarfs, which may lead to underestimated luminosities for the young 
dwarfs.
The bolometric corrections appropriate for the young late M and L dwarfs are not yet available, but a few individual values do exist in the literature. For example, \citet{zapatero10} derive BC$_J$=$1.16 \pm 0.10$ for a young L3 dwarf G196-3B, whereas a field dwarf of the same spectral type
would have BC$_J$=$1.85 \pm 0.14$ \citep{schmidt14}. This difference in BC$_J$ shifts an object by $\sim 0.3$\,dex in the positive y-direction, i.e. enough to fall above the 10 Myr isochrone.
On the other hand, \citet{barman11a, barman11b}, studying two planetary-mass companions to young stars, suggest that 
the effect might be caused by an overestimate of the $T_{\mathrm{eff}}$. They were able to reproduce the observed properties
of the two objects by modeling low-gravity, cloudy atmospheres experiencing non-equilibrium chemistry, and requiring the $T_{\mathrm{eff}}$ that
is several hundred K lower than reported earlier. 
Therefore, the position of the coolest objects in the HR-diagram with respect to the theoretical isochrones can reflect a problem
with the bolometric luminosities and/or the atmosphere models in this $T_{\mathrm{eff}}$ regime. The latter is possibly
due to treatment of dust in various models, as clouds start to form in the atmospheres of cool dwarfs with $T_{\mathrm{eff}} \lesssim 2800\,$K
\citep{helling08, helling14}.



\section{Discussion}
\label{discuss}

\subsection{Chamaeleon-I}
\label{discuss_ChaI}
Based on the survey presented in this work, we can put limits on the number of VLM sources that are missing in the
current census of YSOs in Cha-I. As in other works from the SONYC series, we use the success rates of our 
spectroscopic follow up (defined as a number of confirmed members divided by the number of the spectroscopically surveyed
candidates), to estimate the expected number of members among the remaining photometric candidates.
The listed uncertainties are the 95\% confidence intervals (CI), calculated using the Clopper-Pearson method, 
which is suitable for small-number events, and returns conservative CIs compared to other methods \citep{gehrels86, brown01}.

While until now for this estimate we always considered the entire candidate lists above the completeness limit, in case of Cha-I it might be 
more suitable to divide the candidate list in two, and consider the candidates with $I\leq21$, and those with $21<I<23$ separately. The reason for this is the following.
The candidate list with $I\leq21$ contains 46 objects. We took 16 spectra with VIMOS, and confirmed 7 members.
This gives a success rate of $7/16=44\substack{+26 \\ -24}\%$ (95\% CIs).
Among the candidates with $21<I<23$ that were specifically targeted in this most recent follow-up, we took in total 17 spectra (15 with SofI and 2 with VIMOS), and confirmed only 1 probable VLM members 
of Cha-I (the other red diamond in Figure~\ref{cmd} marks ISO-ChaI-192, whose spectral type is unclear). 
The success rate is therefore $6\substack{+23 \\ -6}\%$, which is significantly lower than for the candidates with  $I\leq21$. 
The difference 
cannot be caused simply by different wavelength domains used in spectroscopy (optical versus NIR), as all but one of the members confirmed with VIMOS
have $T_{\mathrm{eff}}$ below 3500\,K, and would have also been recognized in the NIR. 

From the success rates mentioned above, we can now estimate the number of missing VLM members in our candidate list. 
The number of unconfirmed members among the remaining candidates in a sample can be calculated by subtracting the number of spectra taken by SONYC and
number of objects observed by other groups from the total number of candidates in the sample, multiplied by the success rate.
In the higher brightness bin ($I\leq21$), among
our 46 candidates without SONYC spectra, we find additional 8 objects classified as members, and 4 non-members from \citet{luhman04,luhman07}. Thus, the number of missing VLM objects 
with ($I\leq21$) is $(46-16-8-4)\times0.44\substack{+0.26 \\ -0.24}\ = 8\substack{+5 \\ -4} $.   
If we now consider the $21<I<23$ part of the $Iz$ CMD, it contains 60 
candidates, of which 17 have SONYC spectra. One probable VLM member was confirmed by SONYC, and one additional VLM was confirmed in other surveys. 
We estimate therefore the expected number of members in this bin to be $(60-17-1)\times 0.06\substack{+0.23 \\ -0.06} =  3 \substack{+10 \\ -3}$. 
We expect most of these missing members to be substellar, although some of them might be more massive, embedded members of Cha-I.

With the above estimates in mind, we can say that the census of VLMOs in Cha-I is almost complete, 
down to  $0.004-0.009\,$\solm, for A$_V\leq$5 and age of 2 Myrs.

\subsection{Lupus~3}
\label{discuss_Lupus}


Following the discussion in the Sections~5.3 to 5.5 of \citet{muzic14}, we can update the mass function based on
the follow-up presented in this work. For this we take into account only the candidates above the 
completeness levels of our survey. 
The updated success rate in the ``$IJ$-pm'' sample is therefore $9/50 = 18\substack{+13 \\ -9}\%$, 
while for the photometric-only candidates we have $0/81=0\substack{+5 \\ -0}\%$.
As in the previous section, the number of unconfirmed members among the remaining candidates in a sample can be calculated by subtracting the number of spectra taken by SONYC and
number of objects confirmed by other groups from the total number of candidates in the sample, multiplied by the success rate. Therefore, the 
estimated number of unconfirmed members in our survey is $(53-50)\times 0.18\substack{+0.13 \\ -0.09} + (337-53-81-1)\times0\substack{+0.05 \\ -0} \approx 1 \substack{+10 \\ -1}$. The first 
term in this equation refers to the ``$IJ$-pm'' sample, the second to other (photometric only) candidates, and the quoted error is 
the 95\% CI.
We can repeat the same calculation to get an approximate number of missing BDs above the completeness limit, but
the result is essentially the same as for all VLM objects, i.e. any of the 0 - 11 missing objects could be either 
stellar or substellar.

Based on these numbers, it is evident that very few potential VLM members of Lupus~3 are left to be discovered, at least
down to the completeness limit of our survey, which is equivalent to $0.009-0.02\,$\solm, for A$_V\leq$5. Our survey covers a large
portion of the entire Lupus~3 cloud, and encompasses the entire high-extinction band around the two Herbig Ae/Be members (see Figure~1 in \citealt{muzic14}) where
majority of the known members are located.
 
Comparing Lupus~3 with Cha-I, we note that the success rate of spectroscopic confirmation of VLM candidates is significantly lower in the former, despite having used a homogenous method for selecting candidates.
This can be explained by different contamination rates by background objects along the two lines of sight. 
The galactic longitude of Lupus~3 is $l\sim340^{o}$, which means that we are looking
towards the inner, more densely populated parts of the Galaxy, compared to regions along the line of sight to Chamaeleon ($l\sim297^{o}$). The Besan\c{c}on
 Milky Way stellar population synthesis model \citep{robin03} yields about five times more objects in the direction of Lupus~3, than towards Cha-I, within the same area on the sky and for the same photometry limits. 
 Lupus~3 is at the same time less populated: the current stellar census in this region contains roughly 80 members, while in Cha-I there are $\sim$180. Higher contamination by background sources, combined with a smaller overall population is therefore the most plausible cause for different success rates in our spectroscopic surveys.


\subsection{Distribution of spectral types}
\label{S_sptdistr}

\begin{figure*}
\centering
\resizebox{15cm}{!}{\includegraphics{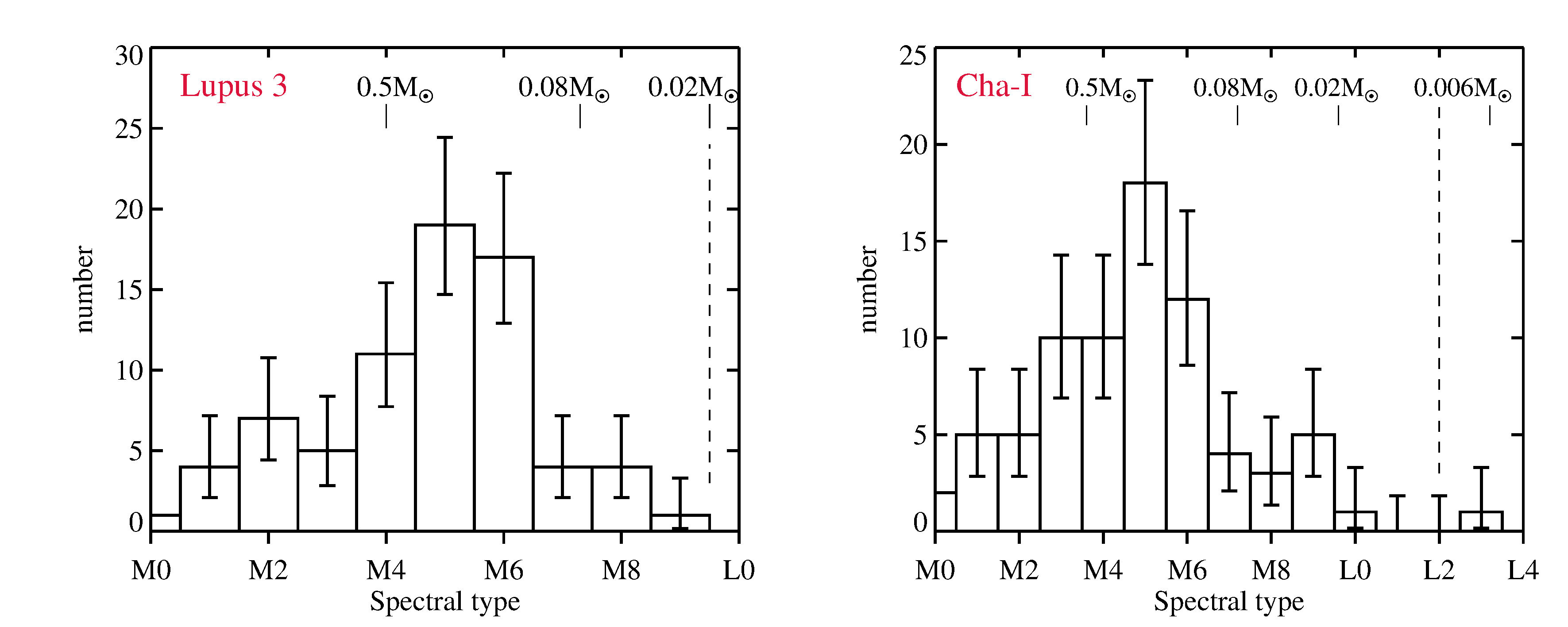}}
\caption{Distribution of spectral types for the VLM population of Lupus$\,$3 (left) and Cha-I (right), within the area of our surveys.
The rough mass limits according to BT-Settl isochrones are shown on top of the plot; the dashed lines mark the completeness limit of the two surveys for A$_V$=5.}
\label{spt_distr}
\end{figure*}

In Figure~\ref{spt_distr} we show the distributions of spectral types for the spectroscopically
confirmed VLM population in Lupus$\,$3 (left panel), and Cha-I (right panel), found within the region covered by our surveys.
The solid error bars are Poissonian
confidence intervals calculated with the method described in \citet{gehrels86}. 
We do not try to correct the histograms for the missing objects estimated in the previous section, because it
is not clear which spectral type they would have. Assuming that they have moderate extinction (A$_V \leq 5$), we could distribute 
them in the substellar bins of the two histograms. However, our previous surveys have revealed a significant number of more embedded
stellar members in the same part of the CMD where those moderately-extincted substellar members are found. In any case, the numbers of missing members are low, 
especially in Lupus~3, and they cannot strongly affect the spectral type distribution.



The overall trend seen in the two histograms is very similar, with a rise in the number of objects at spectral types earlier than M5, a peak at M5, and a drop in the number of objects at later spectral types. A similar behaviour is observed in other star forming regions, e.g. IC 348\citep{luhman07}, 
$\rho$ Oph \citep{ado12}, and NGC~1333 \citep{scholz12b}\footnote{See Figure 13 in \citet{muzic14} for a direct comparison.}. 
The sharp drop in the number of objects at spectral types M7 and later is seen in all these distributions, and is certainly a real feature 
of the IMF, and cannot be attributed to the incompleteness of each survey, especially in Cha-I where we are complete down to $\sim$L3. 
The survey in Lupus~3 is somewhat shallower, but from the analysis in Section~\ref{discuss_Lupus} it is evident that there are very few new
members left to be discovered in this region down to $\sim$L0.
Even if we assume that all the objects missed by our surveys are substellar (i.e. in the bins M7 and later), the drop at late
spectral types would remain.
The two deepest SONYC surveys, in NGC~1333 and Cha-I, both contain very few objects later than M9. The two surveys are both complete down to $\sim$0.005\solm, and reveal that the free-floating objects below D-burning limit are rare. This will be further discussed in the Section~\ref{PMOs}.

\subsection{Star/BD ratio}

\begin{deluxetable}{cccc}
\tabletypesize{\scriptsize}
\tablecaption{Star/BD number ratio}
\tablewidth{0pt}
\tablehead{\multicolumn{4}{c}{\bf Chamaeleon-I}\\[0ex]} 
\tablecolumns{4}
\startdata 
stars & BDs & R$_1$\tablenotemark{a} & R$_2$\tablenotemark{b}\\
\hline
\\
all & all & $2.4-3.8$ & $1.7 - 2.8$\\
$<1$\solm & all & $2.0 - 3.4$ & $1.5 - 2.5$\\
$<1$\solm & $\geq0.03$\solm & $3.0 - 6.1$ & $2.5 - 5.6$\\
\hline
\\
\multicolumn{4}{c}{\bf Lupus~3}\\[1ex]
\hline
\\
stars & BDs & R$_1$\tablenotemark{c} & R$_2$\tablenotemark{b}\\
\hline
\\
all & all & $2.3-3.4$ & $2.0 - 3.0$\\
$<1$\solm & all & $1.7 - 2.7$ & $1.6 - 2.5$\\
$<1$\solm & $\geq0.03$\solm & $3.0 - 6.1$ & $2.9 - 5.8$
\enddata
\label{T_ratio}
\tablenotetext{a}{in the area covered by the Luhman census}
\tablenotetext{b}{in the area of our survey}
\tablenotetext{c}{in the area covered by surveys of \citet{merin08} and \citet{comeron09}}
\end{deluxetable}

To assess the numbers of stars and BDs, we use the approach described in \citet{scholz13}.
In short, by comparing the multi-band photometry\footnote{$IJHK$ photometry complemented with the optical $BVR$ bands where available from public catalogs.} 
with the predictions of the BT-Settl evolutionary models, we derive best-fit mass
and A$_V$ for each object, for the assumed distance of 160$\,$pc and age of 2 Myr for Cha-I, and 200\,pc and 1 Myr for Lupus~3, and the extinction law from \citet{cardelli89} with R$_V=4$. 

For the stellar-substellar mass boundary, we take the value at the solar metallicity, 0.075\solm.
All the objects with estimated masses below 0.065$\,$M$_{\odot}$ are counted as BDs, 
and with masses above 0.085$\,$M$_{\odot}$ as stars. The remaining objects at the border of the substellar regime are once included in the higher mass bin, and then in the lower mass bin, thus resulting in a lower and upper limit of the star/BD ratio.

In Cha-I, we take the census from \citet{luhman04, luhman07}, and complement it with the objects identified in \citet{luhman08,luhmanmuench08}, 
resolved binaries from \citet{daemgen13}, and one VLM object identified here. 
In Table~\ref{T_ratio}, we list various star/BD number ratios, calculated for different mass bins, and for the area of the Luhman census, as well as the area encompassed by our survey. The limits on stellar mass $<1$\solm, and BD mass $\geq0.03$\solm~were chosen to facilitate comparison with our previous works (e.g. \citealt{scholz13}).

The stellar and substellar population of Lupus~3 has not been as thoroughly surveyed by spectroscopy in the past as that of Cha-I. 
If we look at only the spectroscopically confirmed members, it is clear from Fig.~\ref{spt_distr} that the number of objects in the 
M6 bin is larger than the total number of objects in lower mass bins. Since a good fraction of the objects in the M6 bin are at the substellar border, we 
end up with a very large span of values for the star-to-BD ratio. However, Lupus~3 certainly has a larger fraction of candidate sources that still
await a spectroscopic confirmation. For example, the work of \citet{merin08} contains 124 candidates, of which only 46 have been surveyed by spectroscopy. 
We therefore apply a slightly different approach to calculate the star-to-BD ratio in Lupus~3. We use the following (candidate) member lists:
(1) census of spectroscopically confirmed M-type objects from \citet{muzic14}, updated with the newly discovered object SONYC-Lup3-29;
(2) candidates from \citet{merin08}, excluding those already in the census table. The numbers from this sample are multiplied by the 
success rate of the spectroscopic follow-up \citep[36/46;][]{mortier11}, and by the factor 1.25 to account for the fact that the survey is only
sensitive to objects with disks, whose fraction is estimated to be 70-80$\%$;
(3) candidates from \citet{comeron09}, excluding those already in the census table. The numbers from this sample are multiplied by 0.5, which
is the success rate of the spectroscopic follow-up presented in \citet{comeron13};
(4) known members with the spectral type earlier than M, from \citet{comeron08}. The results for the stars-to-BD ratio are shown in Table~\ref{T_ratio}, for the 
area encompassed by our survey, as well as for the larger area covered by \citet{merin08} and \citet{comeron09}. 

The numbers listed in Table~\ref{T_ratio} for Cha-I and Lupus~3 are consistent with each other, and are also generally consistent with the star-to-BD ratios derived for NGC~1333 and IC~348 by \citet{scholz13}.
Clearly, there are a number of uncertainties involved in this calculation, such as the choice of the isochrones and extinction law used to derive masses, or uncertainties in distances to star forming regions. In Lupus~3,
in particular, there could also be some effects of possible incompleteness at the overlap of the different studies.

The numbers in Table~\ref{T_ratio} do not include our estimate for the objects that are ``missing" in our census. In Cha-I, where the stellar population has been thoroughly characterized by spectroscopy in the works of Luhman et al., we estimate that there might still be $11\substack{+11 \\ -5}$ unidentified members
in our selection box. According to the BT-Settl models, at moderate extinctions (A$_V<5$), these objects should be substellar, although some of them might be more embedded stellar members. If these objects would be substellar, the ratio of the star-to-BD ratio R$_2$ could be as low as 1, with the upper limit of 2.3.
In Lupus~3, where we surveyed almost entire population of the photometric and proper motion candidates, the number of missing objects is $0\substack{+10 \\ -0}$.
With this upper 95\% confidence limit, the star-to-BD ratio R$_2$ would be slightly lower: 1.7-2.4 taking into account all stars and BDs, and 1.4-1.9 if we consider only stars below 1\solm. 

To conclude, the star/BD ratio in Cha-I and Lupus~3 is consistent with the findings in other clusters, and we find that for each formed 
BD there are between 1.5 and 6
formed stars. To probe finer differences, one needs a more sophisticated 
analysis and better constraints on fundamental parameters.

\subsection{Initial Mass Function}
\label{s_imf}

\begin{figure}
\centering
\resizebox{8cm}{!}{\includegraphics{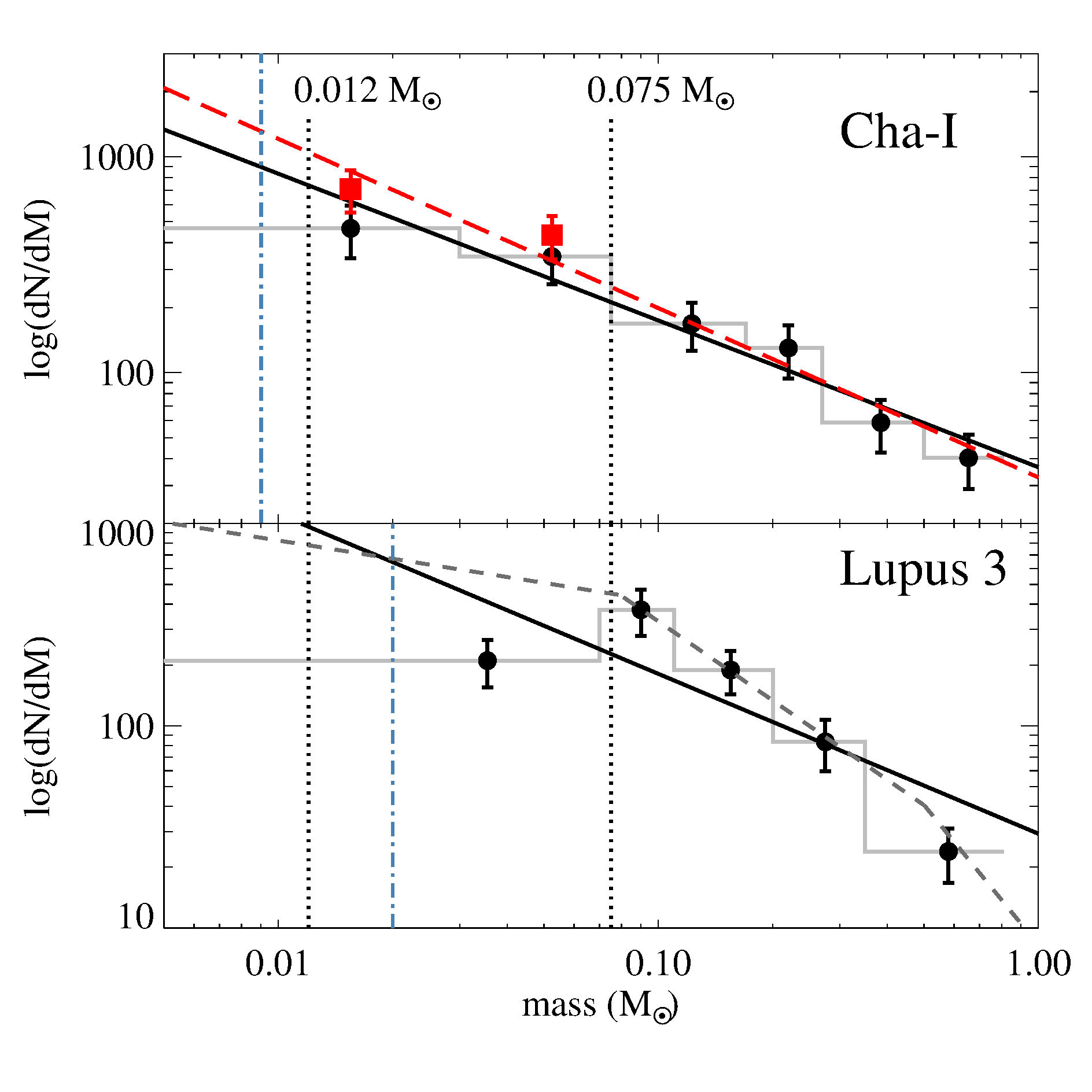}}
\caption{Mass spectrum for low mass stars and BDs in Cha-I (upper panel) and Lupus~3 (lower panel), within the areas of our surveys. 
The bin sizes were chosen to contain similar number of members. The red squares in the Cha-I plot account for the numbers of missing objects (see text). The vertical error bars represent the Poisson uncertainties. The solid black lines show the best fit to the data (black circles), and the red dashed line in the upper panel is the best-fit to the data when the red squares are used instead of the circles at their respective masses. The vertical dotted lines mark the 
approximate locations of H- and D-burning limits, and the dash-dotted lines represent the completeness of the two surveys at A$_V=5$. The grey dashed line in the lower panel is the Kroupa segmented power-law mass function.
 }
\label{imf}
\end{figure}

\begin{figure}
\centering
\resizebox{8cm}{!}{\includegraphics{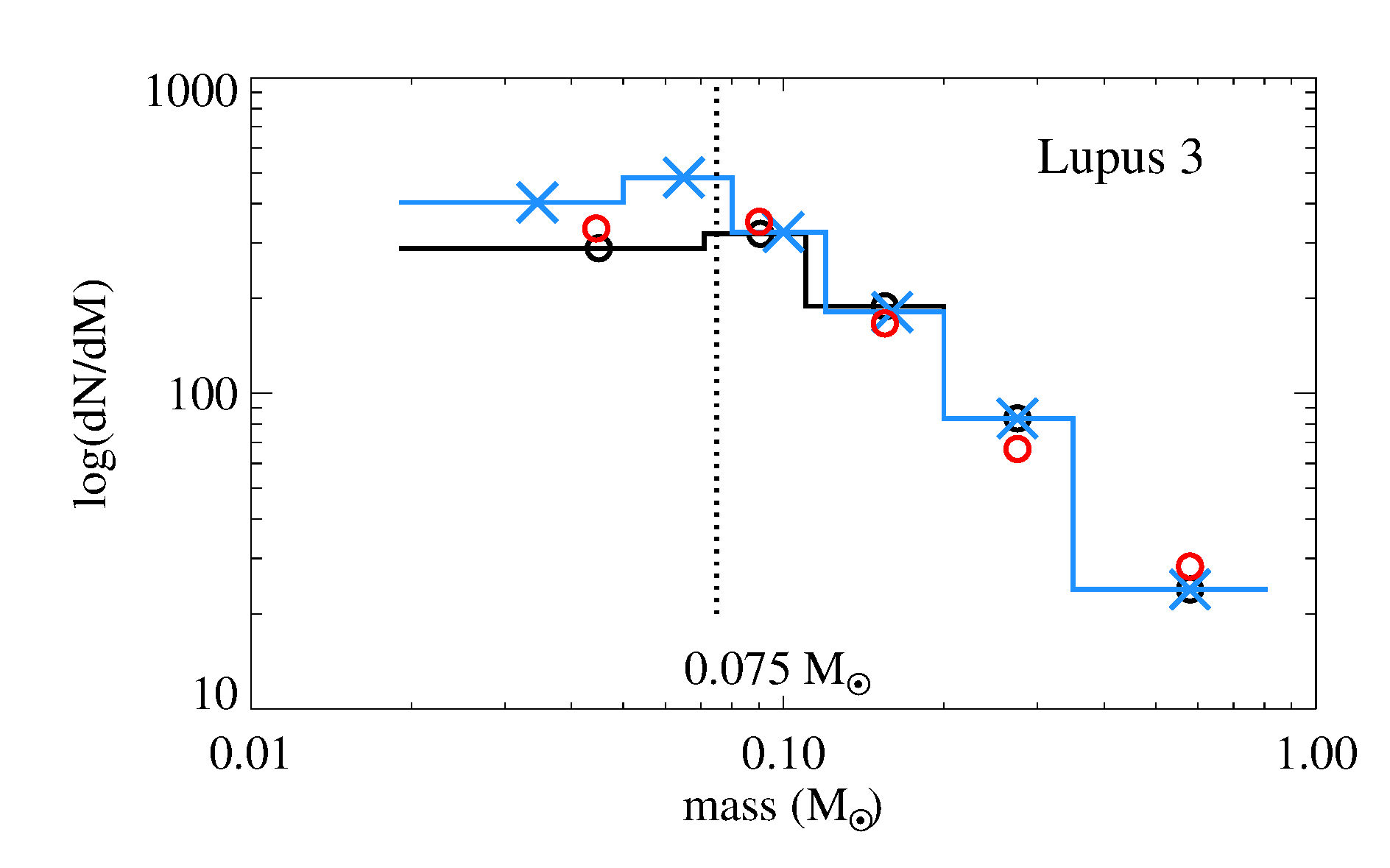}}
\caption{Mass spectrum in Lupus~3. Black circles and histogram show the same data as shown in the lower panel of Figure~\ref{imf}, only
restricted to masses above 0.02\solm, which is our conservative completeness limit at A$_V=5$. 
Red circles show the mass spectrum for the same sample of stars, but with masses calculated after random variation of A$_V$ (see text).
Blue crosses and histogram show the mass spectrum after adding 11 substellar objects with masses above our completeness limit at A$_V$=0 (0.009\solm). The bins are modified to have similar error bars, which are not shown here for clarity, but are similar to those in Figure~\ref{imf}.}
\label{imf_lup}
\end{figure}

With the masses calculated as described in the previous section, we can construct the IMFs in Cha-I and Lupus~3. They
 are shown in Figure~\ref{imf}, where we varied the binsize in order to achieve similar numbers of members in each bin. We choose
to express the IMF in the power-law form, as $dN/dM\propto M^{-\alpha}$.  
The  data needed to reproduce the plot are given in Table~\ref{T_imf}.
In Cha-I the best fit yields $\alpha = 0.68 \pm 0.09$ (black solid line).
We can also account for the number of objects missing in our census, as estimated in Section~\ref{discuss_ChaI}. 
It is not simple to predict the masses that these ``missing'' objects should have, because of the variable extinction in star forming regions, where
more massive embedded stellar members might mix in the color-magnitude space with the less embedded substellar members. However, if we assume the typical
A$_V < 5$, then our missing members should be substellar. 
The objects in the $21<I<23$ bin were included in the lowest-mass bin ($0.001 - 0.03$\solm), as they are expected to have masses below 0.015\,\solm. Objects with $17.5<I<21$
should have spectral type M7 - L0, and we distribute half of those in each of the two lowest-mass (substellar) bins.
With this correction, the best-fit power-law index (dashed red line) becomes $\alpha = 0.78 \pm 0.08$. 
The slope remains essentially unchanged if we put all the missing objects with $17.5<I<21$ to the more massive of the two substellar bins (0.03-0.075\solm).
Despite the caveat mentioned above, this short exercise tells us that even in the (somewhat extreme) case that all the missing members were substellar, the slope of the IMF
would not be strongly affected.

In Lupus~3, we get the best fit single power-law index $\alpha = 0.79 \pm 0.13$ (bottom panel of Figure~\ref{imf}). Here the lowest-mass bin contains all the substellar objects. 
While in Cha-I all the data points seem to be consistent with a single power-law, in Lupus~3 there might be an indication of the flattening of the slope in the substellar regime. 
For a comparison, we show the segmented power-law IMF by \citet{kroupa01}, where the power-law slope changes at 0.5 and 0.08\,\solm. The Kroupa
IMF, normalized to match the middle point of the Lupus~3 IMF at  0.155\solm, is shown by the dashed grey line in the lower panel of Figure~\ref{imf}, with the slopes $\alpha=2.3$ for M$>0.5$\solm, $\alpha=1.3$ for 
$0.08<M<0.5$\solm, and $\alpha=0.3$ for M$<0.08$\solm.
The possible turnover in the power-law mass function in the substellar regime that we observe in Lupus~3, has not been, to our best knowledge, reported in any other spectroscopically confirmed sample of a young cluster.
 One possible exception is the young, sparsely populated association $\eta$ Cha, which beside the population of $\sim$15 stars of spectral type K and M, does not contain any BDs down to 0.015\solm \citep{luhmansteeghs04}. Assuming the star-to-BD ratio of 1.5 -- 6 that we 
find for Lupus~3 and Cha-I, one would expect a population of 3 -- 10 BDs. 
However, a comparison of the mass distributions in $\eta$ Cha, Cha-I, and IC~348 by means of a two-sided Kolmogorov-Smirnov test, does not reveal a significant difference between the three clusters
\citep{luhman09}.

To test the significance of our findings for the Lupus 3 power-law mass function, we first add the upper limit of 11 missing members to the substellar bin. The objects have random masses above our completeness limit for A$_V$=0 (0.009\,\solm), and below the substellar limit. This 
case is extreme, both regarding the number of missing objects (the estimate from Section~\ref{discuss_Lupus} was $1\substack{+10 \\ -1}$), and by the assumption that all the potentially missing members would be substellar, since some, or even all of them could equally be stellar. The result is shown in Figure~\ref{imf_lup}, represented by blue crosses and histogram. 
After adding the missing members, the bins have been redistributed to contain similar numbers of objects. 
Black circles and histogram show the spectroscopic IMF from Figure~\ref{imf}, restricted to masses above our conservative completeness limit at A$_V=5$ (0.02\solm). We see that even after adding the upper limit of the possibly missing objects,
the flattening of the slope remains.

A second suitable test concerns possible influence of the extinction. 
Here, we derived masses and A$_V$ simultaneously by fitting the photometry to the predictions of the evolutionary models. Alternatively, we could have fixed the A$_V$ to the value derived from spectroscopy, or from a single intrinsic color (e.g. $J-K$, as in Section~\ref{S_HRD}). 
Since for the majority of the objects these values do not differ by a large amount, the resulting IMF looks unchanged.
 To further test the impact of the extinction value used to derive masses, we varied
the A$_V$ by a random amount in the interval $\pm 2$\,mag from the spectroscopic value. We have done this 1000 times for each object, and finally adopted the average mass from all the variations. The result is shown in Figure~\ref{imf_lup}, with red open circles. We see that this has a small impact on the shape of the IMF. As mentioned earlier in the paper, members of Lupus~3 typically have low extinction ($\lesssim 5$). In fact, the spectroscopic sample used to plot the IMF contains 98\% of members with A$_V \leq 3$. At A$_V=10$, a 0.07\solm~object would fall right at the completeness limit of our survey, i.e. at the extinctions above 10, we would not be able to detect substellar objects. However, 
we do not expect that many BD could be hiding at high extinctions, because in this case we should have also found some highly embedded
stellar objects. The optical part of our spectroscopic survey would be sensitive to those, but we detect only one such source (SONYC-Lup3-1), whose spectrum and SED suggest a rare geometrical configuration (edge-on disk).      


The single power-law slopes of the mass functions for the two star forming regions derived here agree well, and are also consistent
with typical values found in the literature\footnote{Here we list only the most recently published results, for earlier works
see \citet{scholz12b}.}.
\citet{scholz12b} find $\alpha=0.6\pm0.1$ for low-mass stars and BDs below
0.6\,\solm~in NGC\,1333, whereas \citet{scholz13}, in a slightly modified analysis, report  $\alpha = 0.9-1.0$ for the same cluster.    
In IC348, \citet{ado13} report $\alpha$ of $1.0 \pm 0.3$
and $0.7 \pm 0.4$ for the substellar ($\leq 80 M_{Jup}$) IMF extinction-limited,
and spatially limited to the center of the cluster, respectively. This is consistent with $\alpha = 0.7-0.8$ found in the same cluster by \citet{scholz13}.
\citet{ado13} also find $\alpha= 0.8 \pm 0.4$
 and $0.7 \pm 0.3$ for the substellar regime in $\rho$ Oph, with the extinction limited to A$_V \leq 8$, and A$_V \leq 15$, respectively. For low-mass stars and BDs in $\sigma$~Ori ($0.35 - 0.006$\solm), \citet{penaramirez12} find $\alpha=0.6 \pm 0.2$, 
in agreement with earlier works in the same region \citep{caballero07,bejar11}. 
In Upper Sco, assuming the age of 5-10 Myrs, \citet{lodieu13b,lodieu13c}, report $\alpha=0.5 - 0.7$, for the mass range $\sim$0.03 - 0.005\solm.
\citet{zapatero14a} report $\alpha$ between 0 and 1 for  
the Pleiades cluster, down to $\approx 0.012$\solm.




\begin{deluxetable}{ccc}
\tabletypesize{\scriptsize}
\tablecaption{Values for the mass functions show in Figure~\ref{imf}.}
\tablecolumns{3}
\tablewidth{0pt}
\tablehead{\colhead{Mass interval} &
           \colhead{Mean mass} &
	   \colhead{dN/dM}}
\startdata 
\noalign{\vskip 2mm}
\multicolumn{3}{c}{{\bf Chamaeleon-I}}\\
\hline
\noalign{\vskip 1mm}
0.001 - 0.030 & 0.0155 & 465.52 \\
0.030 - 0.075 & 0.0525 & 344.44 \\
0.075 - 0.170 & 0.1225 & 168.42 \\
0.170 - 0.270 & 0.2200 & 130.00 \\
0.270 - 0.500 & 0.3850 & 58.70 \\
0.500 - 0.810 & 0.6550 & 40.32 \\
\hline
\noalign{\vskip 3mm} 
\multicolumn{3}{c}{{\bf Lupus~3}}\\
\hline
\noalign{\vskip 1mm}
0.001 - 0.070 & 0.0355 & 210.15 \\
0.070 - 0.110 & 0.0900 & 375.00 \\
0.110 - 0.200 & 0.1550 & 188.89 \\
0.200 - 0.350 & 0.2750 & 83.33 \\
0.350 - 0.810 & 0.5800 & 23.91
\enddata
\label{T_imf}
\end{deluxetable}

We now outline the environmental conditions in our two clusters, to try to understand the origin of the difference between the two IMFs 
in the substellar regime.
According to 
\citet{gutermuth09}, with 19 stars$\,$pc$^{-2}$ Lupus~3 has the lowest mean surface density among nearby star forming regions, 
several times lower than e.g. $\rho$\,Oph (120\,pc$^{-2}$) and NGC~1333 (65\,pc$^{-2}$), and also lower than Serpens (36\,pc$^{-2}$), 
Cha-I (30\,pc$^{-2}$), CrA (29\,pc$^{-2}$), or IC~348 (25\,pc$^{-2}$). In terms of peak surface densities, Cha-I is twice as dense as Lupus~3 (800 vs 400\,pc$^{-2}$).
In \citet{scholz13}, we presented evidence for a difference in mass distributions of NGC~1333 and IC~348, under the (plausible) assumption that 
NGC~1333 is closer to us. The denser of the two clusters (NGC~1333) harbors a larger fraction of BDs.
Therefore, frequencies of BDs might depend on stellar densities, which is in line
with predictions for BD formation through gravitational fragmentation of filaments falling into a cluster potential \citep{bonnell08},
dynamical cluster formation (simulations by \citealt{bate12}), or even the disk fragmentation in some cases (see discussion in \citealt{scholz13}).
 
The presence of massive OB stars might influence BD formation in clusters, since BDs could form through photo-evaporation of prestellar cores \citep{whitworth04}. However, nearby star forming regions ($d\lesssim300$\,pc) contain extremely small populations of OB stars (only one B star in Lupus, and three in Cha-I), and their influence on the IMF of these clusters must be negligible.

\subsection{Brown dwarfs with masses below the deuterium-burning limit}
\label{PMOs}
If we assume monotonic continuation of the power law shown in the upper panel of Figure~\ref{imf} below the 
D-burning limit, we would
expect $8 \pm 4$ objects with $0.005 < M<0.015$\solm~in Cha-I. For the slope shown with the red dashed line, which accounts
for the objects possibly missed by our follow-up, we get $12 \pm 5$. 
Our method for mass estimates from photometry yields 6 objects with masses below $\sim0.015\,$\solm~in the area of our survey, and we estimated that there might be $3\substack{+10 \\ -3}$ more in our candidate list. The data are therefore consistent with a monotonic power law
mass spectrum across the D-burning limit, and down to $\sim$0.005\solm. 

In NGC 1333, we find similar monotonic behavior
of the low-mass IMF across the D-burning limit, 
but do not exclude a shallower slope below this limit (i.e. $\alpha \lesssim 0.6$).
In fact, based on more recent information about the mid-infrared colors of young ultracool objects \citep{faherty13}, our
upper limit on the number of the planetary-mass BDs is probably too conservative, and is likely to be lower (4 instead of 8). This 
suggests a possible break in the power-law around D-burning limit.
In $\sigma\,$Ori, \citet{penaramirez12} report the IMF consistent with a smooth power law with $\alpha =0.6 \pm 0.2$ down to 0.004\solm. In the range 0.004 -- 0.003\solm~there are fewer candidate members observed in their deep J-band catalog than one would expect from an extrapolation of the same power-law to these masses, possibly signaling a lower slope in this mass regime.
\citet{lodieu13c} present a deep NIR survey probing the planetary-mass regime in UpSco, complementing their previous work in the same region \citep{lodieu13b}. 
The updated IMF in UpSco, extending down to $\sim0.005$\solm, is consistent with a rising power-law function with $\alpha\sim 0.5$. 
    
In the area of our Cha-I survey, there are in total 94 known members, out of which $\sim$6 appear to have masses below 0.015\solm. Previously we estimated that there might be 0-13 objects missing in this mass regime, and 4-13 among the higher mass BDs. In the current census, the brown dwarfs with masses in the planetary-mass regime seem to comprise only $\sim$6$\%$ of the Cha-I population, whereas their contribution can go up to $\sim$17$\%$ taking into account these estimates. Even in the extreme case, 
the mass budget of these objects can be only about 1$\%$ of the total mass of the cluster.

 
In Lupus~3, our survey is sensitive only to the high-mass tip of the planetary-mass regime, but 
\citet{comeron11} conducted a small-area photometric survey several magnitudes deeper than ours and sensitive to free-floating Jupiters according to the models.
The saturation of their survey matches the completeness limits of ours, and their $3\sigma$ detection limit is at $I_C=25.6\,$mag.
The number of the cool photometric candidates identified in the survey by \citet{comeron11} is consistent with statistical expectation of a background population. 
The survey encompasses a very small part of Lupus~3, about 100 times smaller than the SONYC area, but carefully chosen to match an area known to be
rich in higher mass stellar and substellar members. While a wider area survey sensitive to the lowest masses is still required to confirm the  
result by \citet{comeron11}, it represents a complement to our finding that there are very few substellar members left to be discovered 
down to the mass limits of our survey.
 

To conclude, surveys in various star forming regions clearly show that the IMF below $\sim$0.6\solm, and down to $\sim$0.02\solm~
can be well described by a monotonic power-law with an exponent $\alpha =0.6 - 1.0$. Below the D-burning limit, the slope seems
to be similar, or shallower, but it is certainly not steeper than the one above 0.02\solm. This means that, if the microlensing results \citep{sumi11}
hold, the objects they probe must undergo a different formation path than those studied so far in young clusters. 


\section{Conclusions and summary}
\label{summary}
Here we have presented the further spectroscopic follow-up of the VLM candidate member lists in
Cha-I and Lupus~3, using the NIR spectrograph SofI at the NTT. 
This work is a continuation of our previous efforts described in \citet{muzic11} and \citet{muzic14}. 

In Lupus~3, we obtained 19 new spectra of our high priority photometric and proper motion sample (``$iJ$-pm''), thus 
almost completing the spectroscopic survey of this sample: we have targeted 50 out of 53 candidates above the survey’s completeness limit.
We identified two probable substellar members of Lupus~3, one of which is new and has a NIR spectral type M7.5.
Based on statistical arguments, we conclude that only a very few substellar members are left to be discovered in this region, at least down to the completeness limit of our survey, lying at $i=20.3$, equivalent  $0.008-0.02\,$\solm, for A$_V\leq$5 at 1 Myr. 

The follow-up in Cha-I included 15 candidates with expected masses in the planetary regime, of which only one was 
confirmed as substellar with the spectral type L3 and $T_{\mathrm{eff}}$ of 2200$\,$K. According to the BT-Settl
models, an object at this $T_{\mathrm{eff}}$ should have a mass of $\sim 0.009-0.012$\solm~at the age of 1-2 Myr. 
The comparison of the multi-wavelength photometry of this object with the models yields a mass of $\sim 0.007\,$\solm, at the 
distance of Cha-I and age of 2 Myr.
Based on statistical arguments, we estimate that in the area of our survey, 
$8\substack{+5 \\ -4}$ objects are left to be discovered with $17.5 < I <21$, and $3\substack{+10 \\ -3}\%$ with $21< I <23$. In terms of masses,
the first bin is equivalent $0.005-0.06\,$\solm, and the second one $0.003-0.02\,$\solm, at the age of 1-2 Myr, and A$_V\leq5$. 

The IMF below 1\solm~in both Cha-I and Lupus~3 can be described by a power law $dN/dM\propto M^{-\alpha}$, with the slope
$\alpha \sim 0.7$, in agreement with the results of the surveys in other clusters. 
The same is valid for the star-to-BD ratio, which is found to be between 1.5 and 6.
In Lupus~3, however, we find evidence for a flattening of the slope, or even a possible turnover of the IMF in the power law form in the substellar regime: this region seems to produce less BDs in comparison to majority of other clusters.
The flattening is still 
present even after accounting for the maximum number of objects possibly missed by our survey.

The IMF in Cha-I shows a monotonic behaviour across the 
D-burning limit. The expected total number of substellar objects can be statistically estimated 
from the success rates of our spectroscopic follow-up, and 
this estimate yields numbers consistent with the same power law extending down to our completeness limit, lying at 0.004 -- 0.009 \solm. We estimate that the low-mass members below the D-burning limit contribute of the order $5 - 15\%$ to the total number of Cha-I members, comprising therefore $\lesssim1\%$ of the mass budget of the cluster. 


\acknowledgements{
The authors thank Katelyn Allers, Catarina Alves de Oliveira, and David Lafreni\`ere for sharing their spectra, used for comparison in this work.
This work was co-funded by NSERC grants to RJ.
}

\end{document}